\documentclass[superscriptaddress,twocolumn,nofootinbib,amsmath,amssymb,aps,prd,]{revtex4-2}

\usepackage[colorlinks]{hyperref}
\usepackage{tabularx}
\usepackage{graphicx}
\usepackage{amssymb,amsmath,bm,tensor,braket}
\usepackage[dvipsnames]{xcolor}
\usepackage[varg]{txfonts}
\usepackage{enumerate}
\usepackage{mathtools}
\usepackage{tensor}
\usepackage[capitalize]{cleveref}
\usepackage[utf8]{inputenc}
\usepackage[normalem]{ulem}
\usepackage{dcolumn}%
\usepackage{soul}
\usepackage{subfigure}
\usepackage[normalem]{ulem}
\usepackage{aas_macros}
\usepackage{multirow}
\usepackage{comment}
\usepackage{siunitx}
\usepackage{mathrsfs}
\usepackage[T1]{fontenc}

\begin{document}

\title{Quasinormal Modes of Reissner-Nordström Black Holes in Modified Gravity}

\author{Xia-Tong Tan}

\author{Yu Zhang}
\email{zhangyu_128@126.com (corresponding author)}

\author{Zhi-Hong Nie}

\affiliation{Faculty of Science, Kunming University of Science and Technology, Kunming 650500, China}

\begin{abstract}
In the framework of Scalar-Tensor-Vector Gravity (STVG), this paper investigates the quasinormal modes and time-domain evolution of Reissner-Nordström Modified Gravity (MOG) black holes under massless scalar, electromagnetic, and gravitational perturbations. Quasinormal frequencies are calculated by extending the 3rd- to 6th-order WKB method and combining it with Padé approximation to the 13th order, while complete time-domain waveforms are obtained using a finite-difference method based on light-cone coordinates. The results indicate that increasing the MOG parameter \(\alpha\) raises the effective mass of the black holes, leading to a decrease in the damping rate \(|\omega_I|\) and a slower ringdown decay. An increase in charge \(Q\) produces a similar effect, with the damping approaching zero in the near-extremal limit. Regarding the ringdown decay times for the three types of perturbations, the gravitational field has the longest duration, the electromagnetic field is intermediate, and the scalar field is the shortest. The oscillations and decay observed in the time-domain waveforms are qualitatively consistent with the frequency-domain results, and the power-law exponent governing the late-time tail aligns with that of asymptotically flat spacetimes. This study provides a systematic theoretical reference for testing modified gravity theories using black hole spectroscopy.
\end{abstract}

\maketitle

\section {Introduction}
Gravitational-wave astronomy has ushered in the era of black hole spectroscopy~\cite{Berti_2009}. Black hole quasinormal modes (QNM), characteristic oscillations of spacetime itself, depend solely on the black hole's background parameters rather than the initial perturbations, earning them the title of black hole ``fingerprints''~\cite{Berti_2009,Kokkotas1999,RevModPhys.83.793}. The direct detection of post-merger ringdown signals by ground-based observatories like LIGO and Virgo has transformed QNM into exceptionally powerful probes for testing general relativity and investigating modified gravity effects in the strong-field regime~\cite{Berti_2009,PhysRevLett.116.221101,PhysRevD.94.084002}. Since any theory deviating from general relativity could leave observable imprints on the QNM spectrum, the systematic study of black hole QNM properties within various modified gravity frameworks holds significant theoretical value and potential observational importance~\cite{Cardoso2019}.

Among the various modified gravity schemes, Scalar-Tensor-Vector Gravity (STVG)~\cite{Moffat_2006}, often referred to as MOG, has garnered significant attention for its ability to explain galaxy rotation curves, galaxy cluster dynamics, and certain cosmological observations without the need for dark matter~\cite{Brownstein_2006,10.1093/mnras/stt1670,GREEN2019100323}. The theory's core mechanism involves dynamically modifying the effective gravitational constant via a scalar field and introducing a vector field that couples directly to matter. This generates an effective repulsive force in weak-field regions, thereby modifying Newtonian gravity. In strong-field regions, STVG yields black hole solutions that differ from those of General Relativity~\cite{Moffat2015,PhysRevD.110.024042,Endlich2017,Yunes2013,Ishak2019}. When electric charge is taken into account, the static spherically symmetric Reissner-Nordström  MOG (RN-MOG) black hole emerges as a significant spacetime configuration naturally predicted by the theory. Its metric structure depends not only on the black hole's mass and charge but also sensitively on a dimensionless MOG parameter $\alpha$, which characterizes the deviation from General Relativity. As $\alpha \to 0$, the spacetime smoothly reverts to the standard Reissner-Nordström solution. As both $\alpha \to 0$ and $Q \to 0$, it smoothly reverts to the standard Schwarzschild solution. This modification of the background metric introduces novel characteristics regarding the propagation and decay of perturbations around the black holes, thereby enabling direct testing of STVG through the analysis of quasinormal modes.

Massless scalar fields, electromagnetic fields (vector fields), and gravitational fields represent the three most fundamental and significant types of perturbations for probing the spacetime geometry of black holes~\cite{Berti_2009,RevModPhys.83.793}. Within the framework of the Reissner-Nordström black holes in general relativity, scalar, electromagnetic, and gravitational perturbations can all be reduced to one-dimensional Schrödinger-like equations characterized by effective potentials. The corresponding quasinormal frequencies have been extensively studied. In the STVG framework, modifications to the background metric introduce an additional parameter $\alpha$ into these effective potentials, thereby offering an opportunity to directly test modified gravity theories using quasinormal modes~\cite{Moffat2015}. Systematically calculating the quasinormal frequencies and late-time time-domain evolution of RN-MOG black holes under these perturbations not only reveals how MOG parameters influence the black holes' response characteristics but also establishes a clear theoretical template for future constraints on STVG theory via gravitational wave or electromagnetic observations.

The structure of this paper is as follows. In Sec.~\ref{chapter2}, we introduced RN-MOG Black Hole Background. In Sec.~\ref{chapter3}, we derive the wave equations governing massless scalar, electromagnetic, and gravitational perturbations in the spacetime of RN-MOG black holes, and present their respective effective potentials. In Sec.~\ref{chapter4}, we calculate the quasinormal frequencies for these three types of perturbations using both the higher-order WKB method and Padé approximation, and analyze in detail how the frequencies vary with the MOG parameter $\alpha$, charge $Q$, and angular quantum number $l$. In Sec.~\ref{chapter5}, we employ the finite-difference method to numerically evolve the perturbation equations, obtaining complete time-domain waveforms that span from early-stage quasinormal oscillations to late-stage power-law tails. Finally, Sec.~\ref{chapter6} presents the conclusions. By systematically comparing the quasinormal modes characteristics of MOG black holes with those of general relativistic black holes, this paper aims to reveal the specific impact of STVG corrections on black hole dynamical stability and response spectra, thereby providing a theoretical basis for testing modified gravity theories through future astronomical observations.

Throughout this paper, we use geometric units with $c = G = 1$, set the mass parameter to $M = 1$, and adopt the metric signature $(-, +, +, +)$.

\section{RN-MOG Black Hole Background}\label{chapter2}

\subsection{Scalar-Tensor-Vector Gravity (STVG) theory}
STVG theory is a covariant modified gravity theory aimed at replacing the hypotheses of dark matter and dark energy. Its action can be written as~\cite{Moffat_2006,10.1093/mnras/stt1670}:
\begin{equation}
	S = S_{\rm GR} + S_{\phi} + S_{V} + S_{M},
\end{equation}
where $S_{\rm GR}$ is the Einstein-Hilbert action, $S_{\phi}$ comprises the kinetic and potential terms for the scalar field $G$ (or equivalently $\alpha$), $S_{V}$ contains the kinetic and mass terms for the vector field $\phi^\mu$, and $S_{M}$ represents the coupling of matter fields to the gravitational and vector fields. The central features of the theory are:
\begin{itemize}
	\item The scalar field $\alpha = (G - G_N)/G_N$ describes the spacetime variation of the effective gravitational constant, where $G_N$ is Newton's constant.
	\item The vector field $\phi^\mu$ generates a repulsive-like force through its coupling to the matter current, with the coupling strength regulated by $\alpha$.
	\item In the weak-field, low-velocity limit, the theory yields a modified Newtonian potential: $V(r) = -G_N(1+\alpha)M/r + \alpha G_N M e^{-r/r_0}/r$, where $r_0$ is related to the mass of the vector field. On galactic scales, the repulsive term can explain galaxy rotation curves without invoking dark matter.
\end{itemize}
In the strong-field regime, the mass of the vector field becomes negligible and the scalar field $\alpha$ approaches a constant. the theory then reduces to Einstein-Maxwell-type equations with a constant parameter $\alpha$, supplemented by a vector field coupled to matter.

\subsection{Static spherically symmetric RN-MOG black hole solution}

\begin{figure*}[tb]
	\centering
	\includegraphics[angle=0, width=0.8\textwidth]{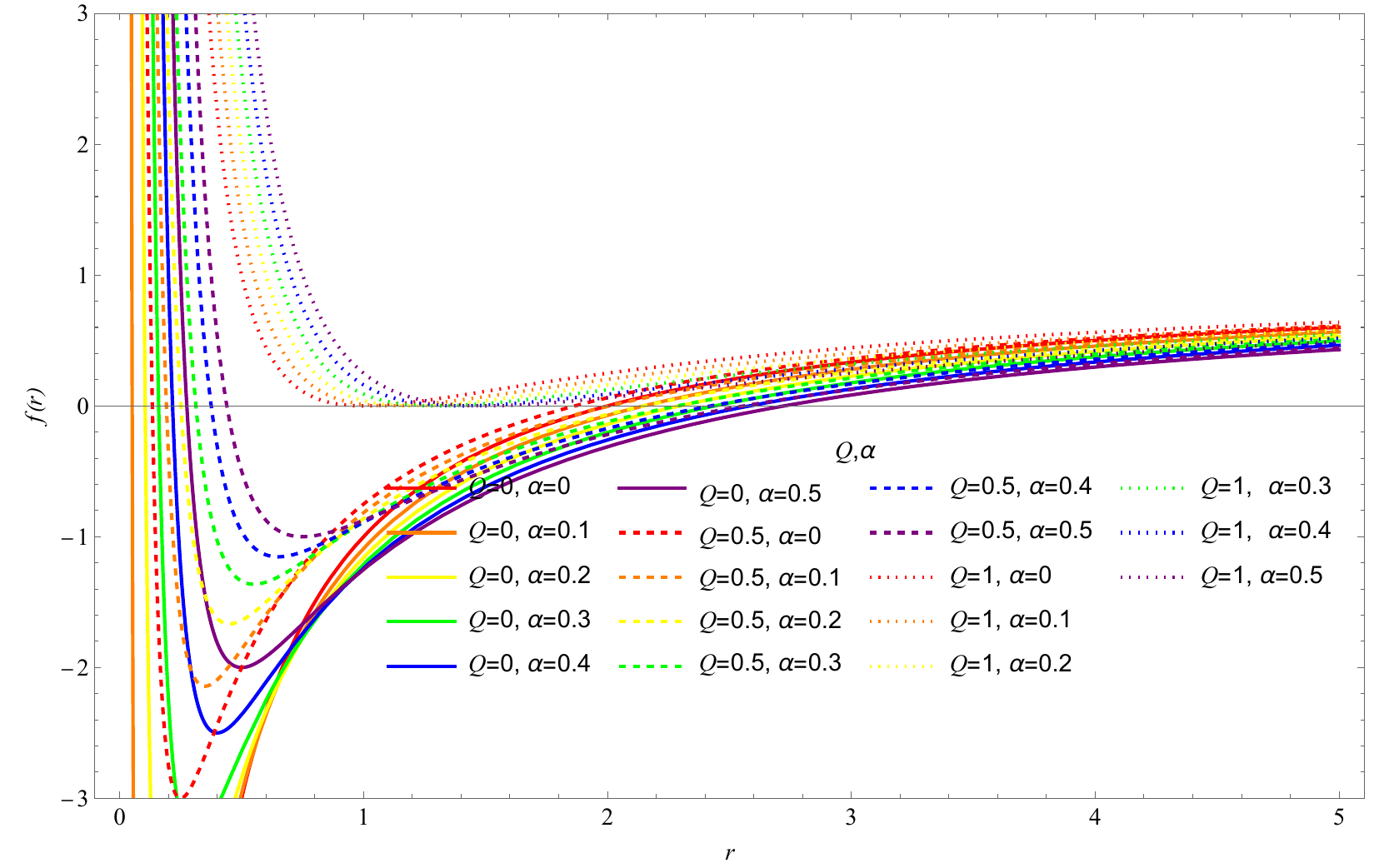}
	\caption{Radial behavior of the metric function $f(r)$ for different parameters.}
	\label{f1}
\end{figure*}

When the mass of the vector field is neglected and the scalar field $\alpha$ is assumed to be constant, the STVG field equations admit a static, spherically symmetric, charged black hole solution, namely, the Reissner-Nordström MOG black hole solution. Its line element is given by~\cite{Nishonov2025}:
\begin{equation}
	ds^2 = -f(r) dt^2 + \frac{1}{f(r)} dr^2 + r^2 (d\theta^2 + \sin^2\theta d\varphi^2),
\end{equation}
where
\begin{equation}
	f(r) = 1 - \frac{2(1+\alpha)M}{r} + (1+\alpha)\frac{(\alpha M^2+Q^2 )}{r^2}.
\end{equation}

Here, $M$ is the mass of the black hole, $Q$ is the charge, and $\alpha$ is a dimensionless MOG parameter ($\alpha \ge 0$) representing the degree of deviation from General Relativity. As $\alpha \to 0$, $(1+\alpha)M \to M$ and $(1+\alpha)(\alpha M^2+Q^2 ) \to Q^2$, and the spacetime smoothly reverts to the standard RN black hole. When $\alpha \to 0$ and $Q \to 0$, it reduces to the Schwarzschild black hole.

\subsection{Horizon Structure and Parameter Space}
The event horizons of the black hole are determined by the roots of $f(r)=0$, namely
\begin{equation}
	r_{\pm} = (1+\alpha)M \pm \sqrt{(1+\alpha)^2M^2 - (1+\alpha)(\alpha M^2+Q^2 )}.
\end{equation}

The condition for the existence of event horizons is
\begin{equation}
	(1+\alpha)^2M^2 \ge (1+\alpha)(\alpha M^2+Q^2 ),
\end{equation}
that is,

\begin{equation}
	M^2 \ge Q^2.
\end{equation}

As shown in Fig.~\ref{f1}, the RN-MOG solution exhibits horizon properties that depend on the parameters. If $\alpha$ and $Q$ are sufficiently large, the metric function $f(r)$ does not cross the zero line and remains positive, indicating the disappearance of the event horizon. Consequently, the existence of an event horizon constrains the parameters to $\alpha \in [0, 0.5]$ and $Q \in [0, 1]$~\cite{Nishonov2025}. Furthermore, at radii far exceeding the horizon, the spacetime differs significantly from the Schwarzschild spacetime, implying that weak-field tests provide limited constraints on these modifications~\cite{Yue2025}. Therefore, assessing the effects of $\alpha$ and $Q$ requires probing the strong-gravity regime~\cite{universe7010009}. However, $\alpha$ alters the relative positions of the inner and outer horizons:
\begin{itemize}
	\item For given values of $M$ and $Q$, increasing $\alpha$ causes both $(1+\alpha)M$ and $(1+\alpha)(\alpha M^2+Q^2 )$ to increase. However, the radius of the outer horizon may either increase or decrease, depending on the magnitude of $Q$.
	\item The horizon of an extremal black hole ($M^2=Q^2$) is located at $r_{\rm ext} = (1+\alpha)M$~\cite{Nishonov2025}. Compared to the GR case where $r_{\rm ext}=M$, the MOG extremal black hole is larger in scale.
\end{itemize}
Such non-trivial horizon modifications directly affect the propagation of perturbations near the black hole, thereby leaving a unique imprint of the MOG parameter $\alpha$ on the quasinormal mode spectrum. In the next section, we will formulate the perturbation equations and corresponding effective potential for various massless test fields within this framework.

\section{The perturbation equation and effective potential}\label{chapter3}
Since the RN-MOG black hole is a charged black hole, there are no purely electromagnetic or purely gravitational oscillations~\cite{PhysRevD.96.024011}. Given that the axial and polar perturbations are isospectral, we only consider the effective potential for the axial perturbations.
Therefore, the following analysis is conducted under the test-field approximation, meaning the perturbation fields do not affect the background geometry, and considering only the axial part of the metric for both electromagnetic and gravitational perturbations, while neglecting perturbations in the additional MOG degrees of freedom. The perturbation equations and effective potentials are derived for massless fields of spin $s=0$ (scalar field), $s=1$ (electromagnetic/vector field), and $s=2$ (gravitational field), respectively.

\subsection{Scalar field (\(s=0\))}
A massless scalar field \(\Phi\) satisfies the Klein-Gordon equation~\cite{PhysRevD.102.084061,Betschart_2004}:
\begin{equation}
	\Box\Phi\equiv\frac{1}{\sqrt{-g}}\partial_\mu\big(\sqrt{-g}g^{\mu\nu}\partial_\nu\Phi\big)=0.
\end{equation}
Given the spherically symmetric background, we separate variables as
\begin{equation}
	\Phi(t,r,\theta,\varphi)=\frac{1}{r}\Psi(t,r)Y_{l m}(\theta,\varphi),
\end{equation}
where \(Y_{l m}\) are the spherical harmonics. Introducing the tortoise coordinate \(dr_* = dr/f(r)\) and assuming a time-harmonic form \(\Psi(t,r)=e^{-i\omega t}\psi(r)\), the radial equation becomes
\begin{equation}
	\frac{d^2\psi}{dr_*^2}+\big[\omega^2-V_s(r)\big]\psi=0,
\end{equation}
where the effective potential is
\begin{equation}
	V_s(r)=f(r)\left[\frac{l(l+1)}{r^2}+\frac{f'(r)}{r}\right].
\end{equation}

\begin{figure*}[tb]
	\centering	
	\begin{minipage}{0.48\textwidth}
		\centering
		\includegraphics[width=\linewidth]{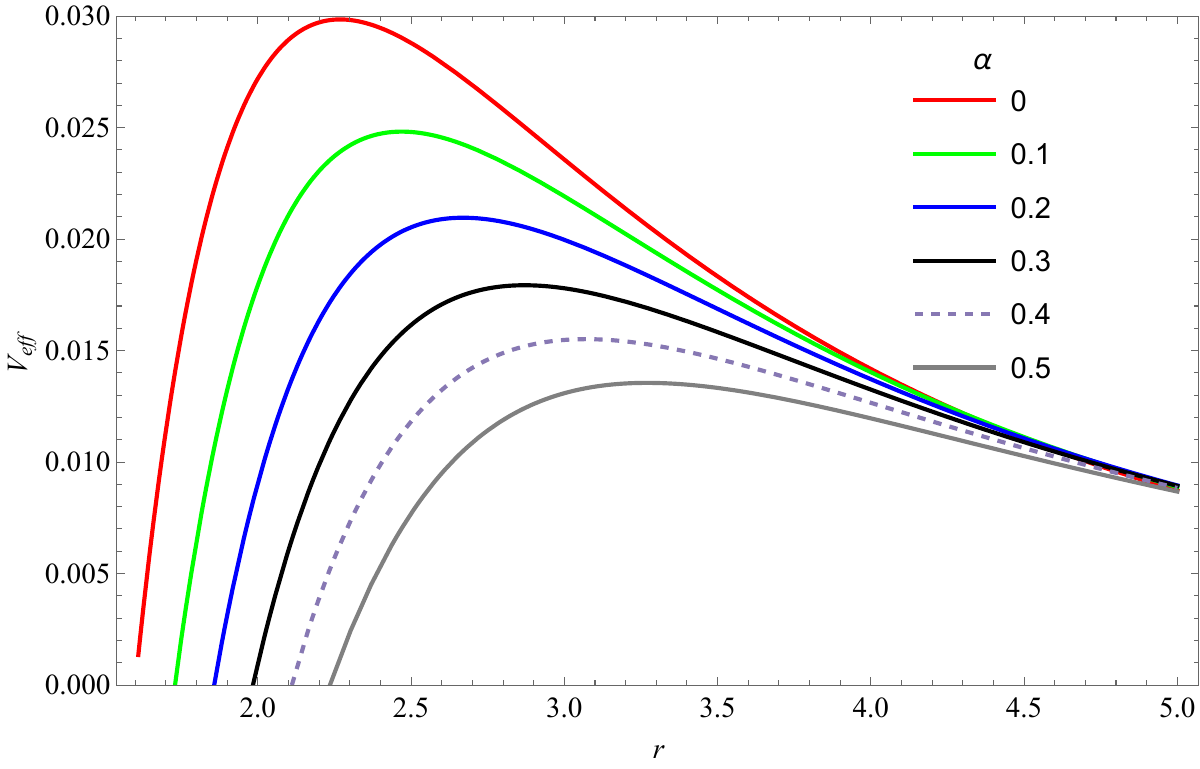}		
	\end{minipage}
	\hfill
	\begin{minipage}{0.48\textwidth}
		\centering
		\includegraphics[width=\linewidth]{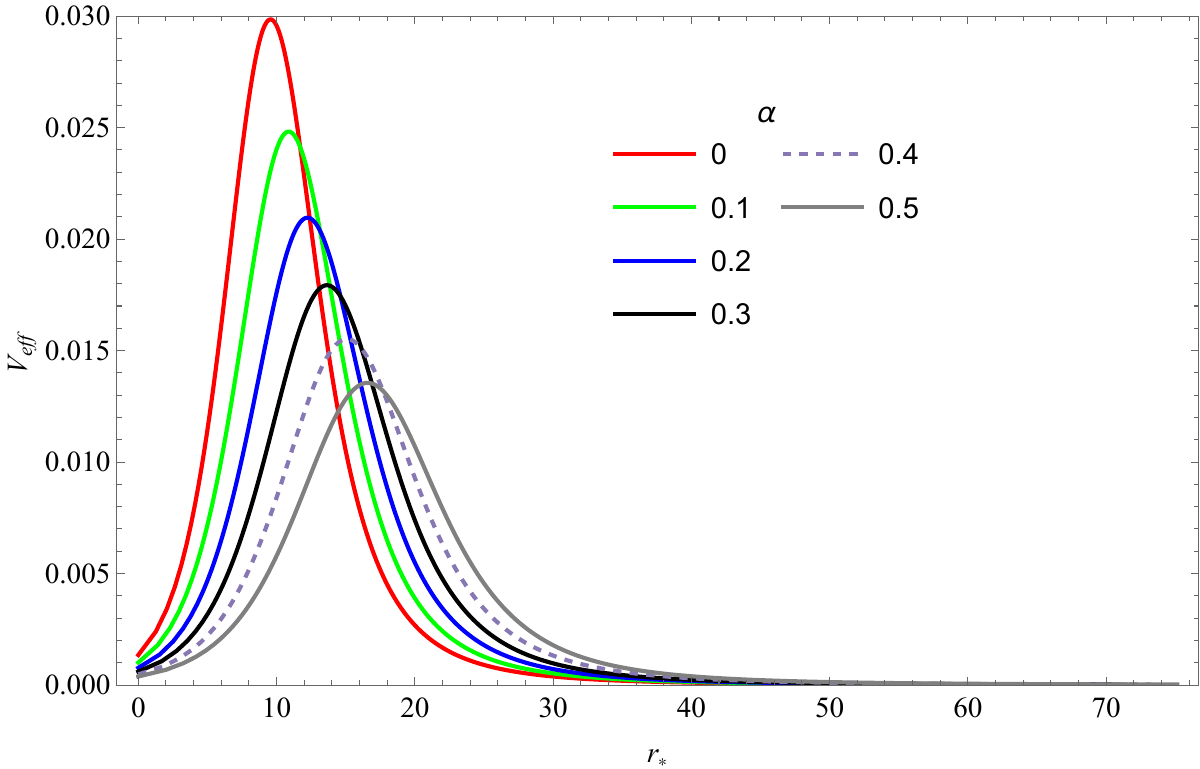}
	\end{minipage}
	    \caption{Effective potential $V_{\rm{eff}}$ for scalar perturbations ($s=0$) of RN-MOG black holes as a function of the MOG parameter $\alpha$, for fixed $Q=0.8$ and $l=0$. Left panel: standard coordinates. right panel: tortoise coordinates.}
	\label{f2}
\end{figure*}

\begin{figure*}[tb]
	\centering
	\begin{minipage}{0.48\textwidth}
		\centering
	    \includegraphics[width=\linewidth]{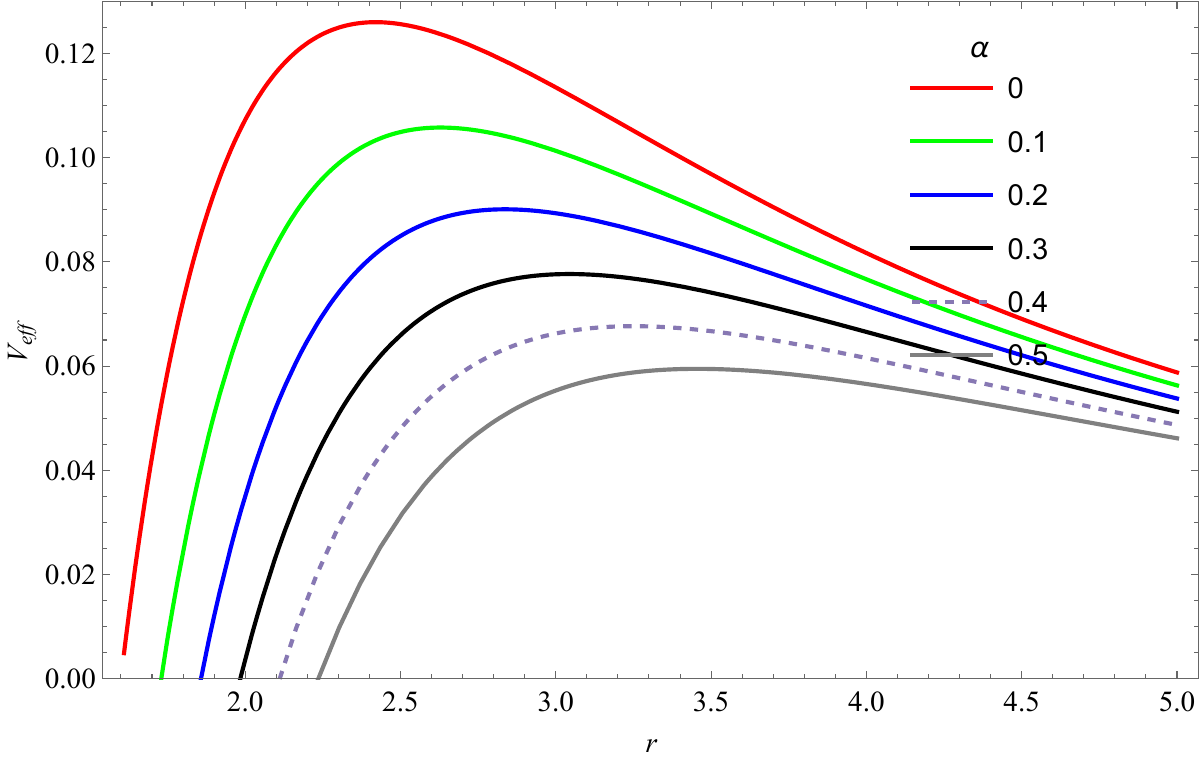}
    \end{minipage}
	\hfill
	\begin{minipage}{0.48\textwidth}
		\centering
	    \includegraphics[width=\linewidth]{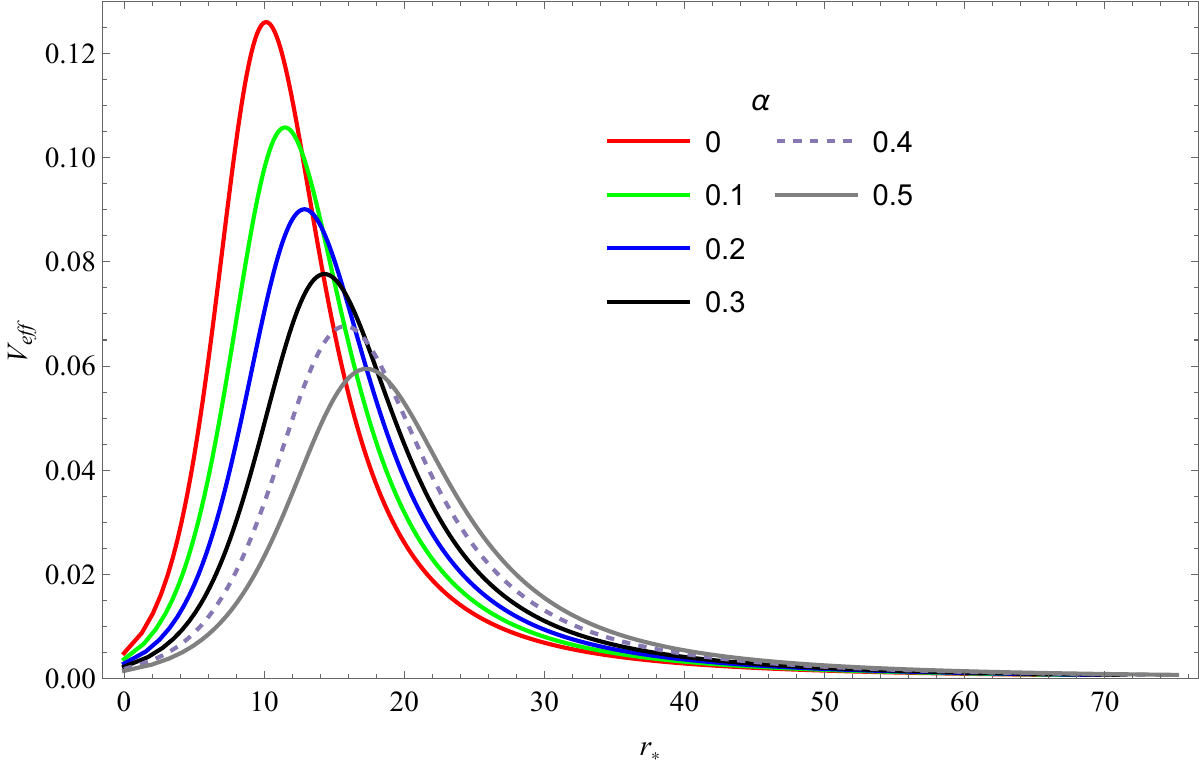}
    \end{minipage}
	\caption{Effective potential $V_{\rm{eff}}$ for scalar perturbations ($s=0$) of RN-MOG black holes as a function of the MOG parameter $\alpha$, for fixed $Q=0.8$ and $l=1$. Left panel: standard coordinates. right panel: tortoise coordinates.}
	\label{f3}
\end{figure*}

We fix the parameter at $Q=0.8$ and plot the effective potential for $l=0$ and $l=1$ against the MOG parameter $\alpha$ in both standard and tortoise coordinates, as shown in Figs.~\ref{f2} and \ref{f3}. We observe that the peak of the effective potential increases with the angular quantum number $l$. Under the influence of the MOG parameter $\alpha$, as it increases, the position of the potential peak shifts to the right, while the height of the peak decreases. Since the calculation of quasinormal modes (QNM) via the WKB method relies on the peak of the effective potential, any shift in the peak's position inevitably affects the behavior of the QNM. When $\alpha=0$, the system reverts to the standard Reissner-Nordström (RN) black hole, represented by the red line in the figures.

\begin{figure*}[tb]
	\centering
	\begin{minipage}{0.48\textwidth}
		\centering
		\includegraphics[width=\linewidth]{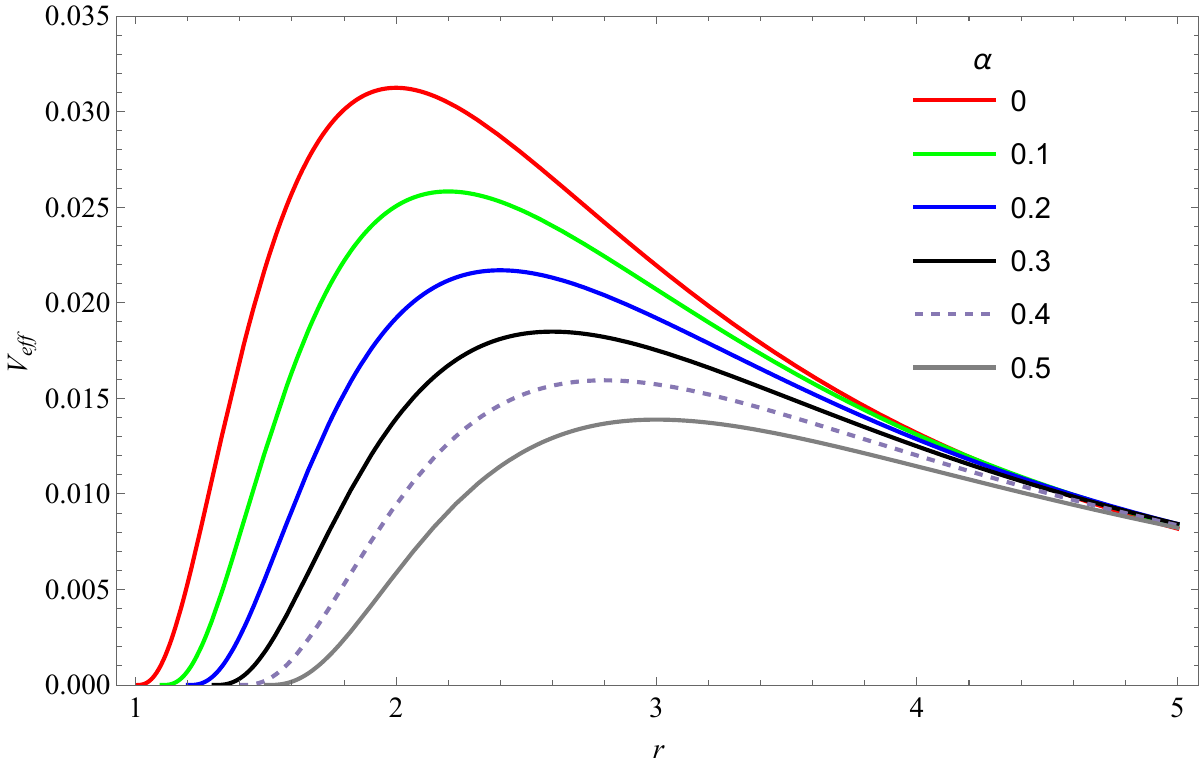}
	\end{minipage}
	\hfill
	\begin{minipage}{0.48\textwidth}
		\centering
		\includegraphics[width=\linewidth]{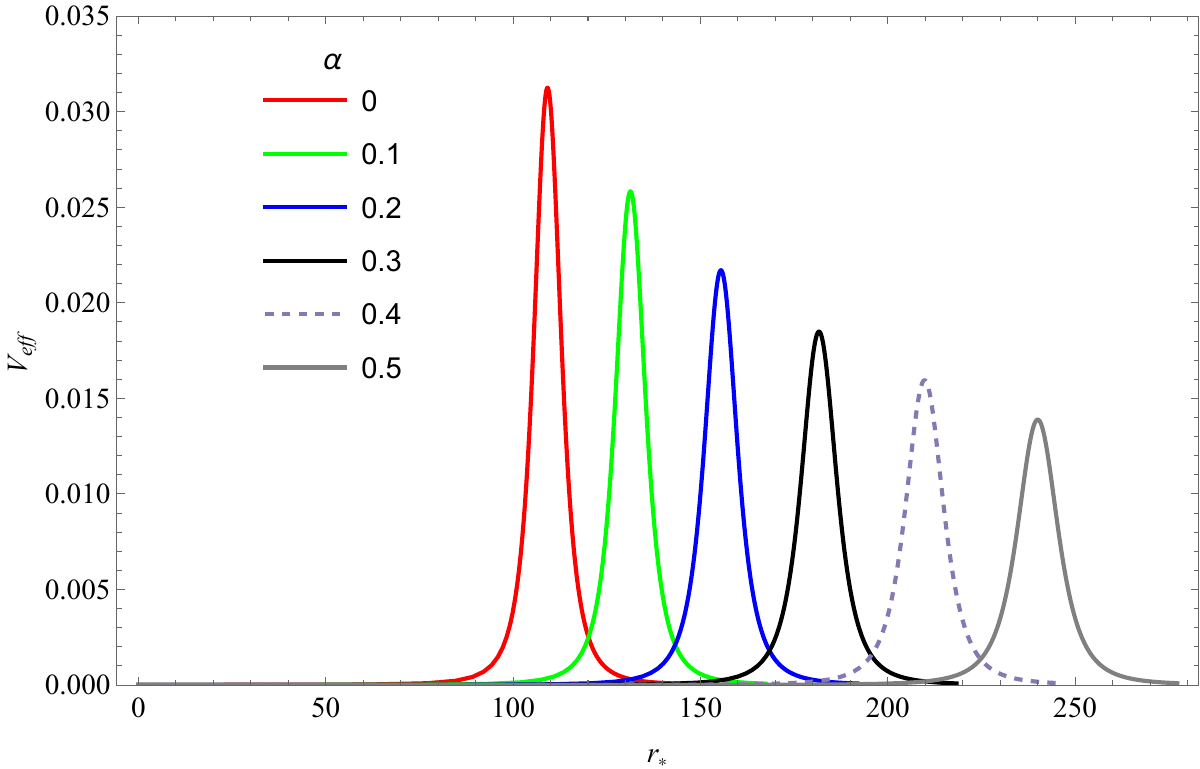}
	\end{minipage}
	\caption{Effective potential $V_{\rm{eff}}$ for scalar perturbations ($s=0$) of RN-MOG black holes as a function of the MOG parameter $\alpha$, for fixed $Q=1$ and $l=0$. Left panel: standard coordinates. right panel: tortoise coordinates.}
	\label{f4}
\end{figure*}

\begin{figure*}[tb]
	\centering
	\begin{minipage}{0.48\textwidth}
		\centering
		\includegraphics[width=\linewidth]{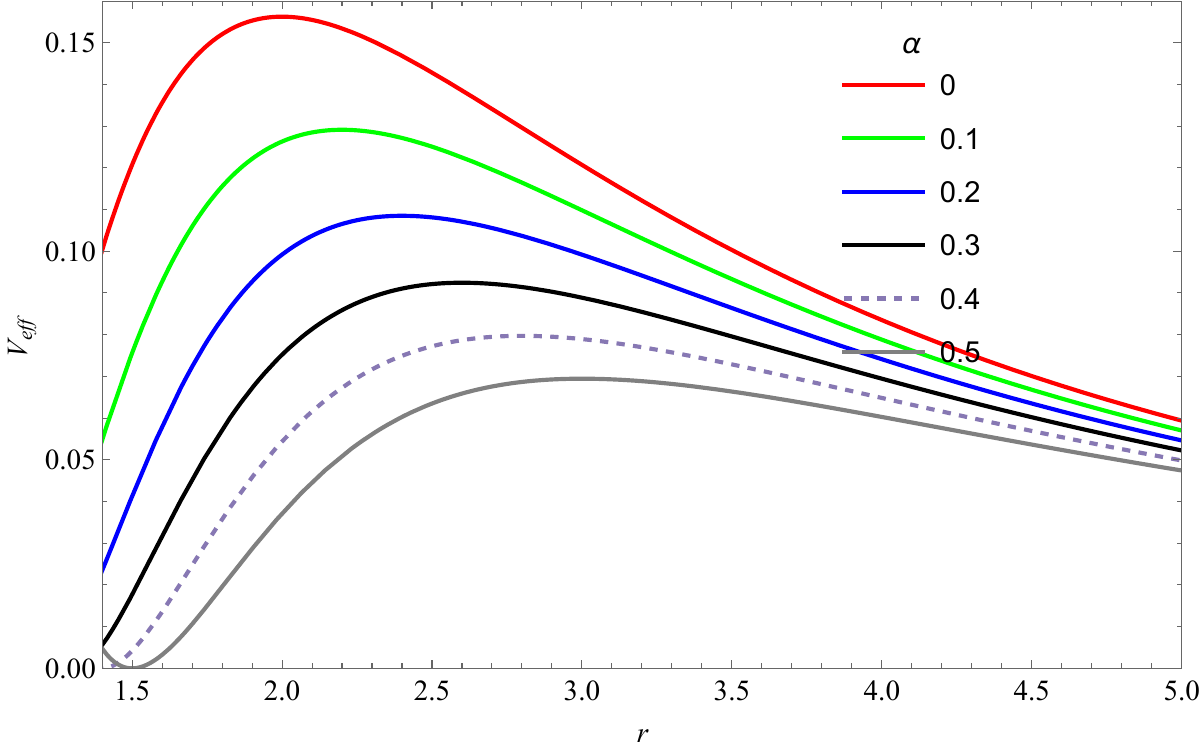}
	\end{minipage}
	\hfill
	\begin{minipage}{0.48\textwidth}
		\centering
		\includegraphics[width=\linewidth]{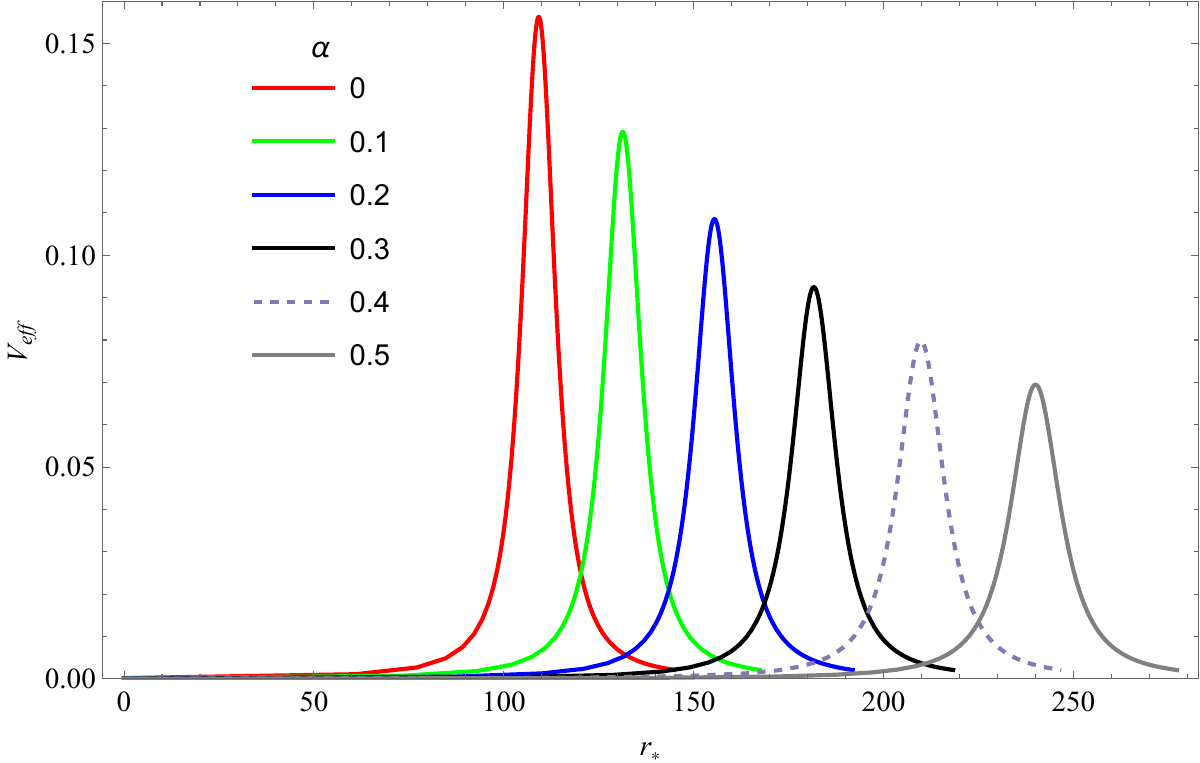}
	\end{minipage}
	\caption{Effective potential $V_{\rm{eff}}$ for scalar perturbations ($s=0$) of RN-MOG black holes as a function of the MOG parameter $\alpha$, for fixed $Q=1$ and $l=1$. Left panel: standard coordinates. right panel: tortoise coordinates.}
	\label{f5}
\end{figure*}

With $Q$ fixed at 1, we plot the effective potential for $l=0$ and $l=1$ against the MOG parameter $\alpha$ in both standard and tortoise coordinates, as shown in Figs.~\ref{f4} and \ref{f5}. We observe that the peak of the effective potential increases with the angular quantum number $l$. Under the influence of the MOG parameter $\alpha$, the peak position shifts to the right as $\alpha$ increases, while the peak height decreases. By comparing Figs.~\ref{f2}--\ref{f5}, we find that the peak of the effective potential increases with the charge $Q$. When $\alpha=0$, the system reverts to the standard RN black hole, as indicated by the red lines in the figures.

\subsection{Electromagnetic field (vector field, \(s=1\))}
The electromagnetic field is described by Maxwell's equations~\cite{Betschart_2004,PhysRevD.100.044017}:
\begin{equation}
	\nabla_\nu F^{\mu\nu}=0,\qquad F_{\mu\nu}=\partial_\mu A_\nu-\partial_\nu A_\mu.
\end{equation}
In a spherically symmetric background, $A_\mu$ is expanded using vector spherical harmonics. In the gauge \(A_0=0\), we separate the variables,
\begin{equation}
	A_\varphi \sim e^{-i\omega t}\psi_{\rm em}(r)\sin\theta\,\partial_\theta Y_{l0},
\end{equation}
substitution into Maxwell's equations and the use of tortoise coordinates yield the master equation
\begin{equation}
	\frac{d^2\psi_{\rm em}}{dr_*^2}+\big[\omega^2-V_{\rm em}(r)\big]\psi_{\rm em}=0,
\end{equation}
where the effective potential is
\begin{equation}
	V_{\rm em}(r)=f(r)\,\frac{l(l+1)}{r^2}.
\end{equation}

\subsection{Gravitational field ($s=2$)}
The metric perturbations $h_{\mu\nu}$ satisfy the linearized Einstein equations and can be decomposed into axial and polar modes against a spherically symmetric background. Axial perturbations are treated using the Regge-Wheeler method, while polar perturbations are handled via the Zerilli method~\cite{PhysRevD.96.024036}. For gravitational perturbations ($s=2$), although both axial and polar modes can be reduced to one-dimensional wave equations of the Schrödinger type, their effective potential generally differ (known as the Regge-Wheeler potential and the Zerilli potential, respectively). These potentials are related by the Chandrasekhar transformation (rendering them isospectral)~\cite{Nagar_2005}and share the same quasinormal frequencies. Therefore, in this paper, we adopt the Regge-Wheeler gauge, and the non-zero perturbation components can be expressed as:
\begin{equation}
	\begin{split}
		h_{t\varphi}=e^{-i\omega t}h_0(r)\sin\theta\,\partial_\theta P_l(\cos\theta),\\
		h_{r\varphi}=e^{-i\omega t}h_1(r)\sin\theta\,\partial_\theta P_l(\cos\theta).
	\end{split}	
\end{equation}
Introducing the gauge-invariant quantity
\begin{equation}
	\Psi_{\rm RW}(r)=\frac{f(r)}{r}\,h_1(r),
\end{equation}
the linearized field equations can be reduced to a Schrödinger-type equation
\begin{equation}
	\frac{d^2\Psi_{\rm RW}}{dr_*^2}+\big[\omega^2-V_{\rm grav}(r)\big]\Psi_{\rm RW}=0.
\end{equation}
The final form of the effective potential is
\begin{align}
	V_{\rm grav}(r) = f(r) \Bigg[ & \frac{l(l+1)}{r^2} - \frac{6(1+\alpha)M}{r^3} \nonumber\\
	& + \frac{4(1+\alpha)(\alpha M^2+Q^2)}{r^4} \Bigg],\label{eff} 
\end{align}
where the effective potential $V_{\rm grav}(r)$ is an approximate result obtained under the test-field approximation \cite{PhysRevD.59.084003}, since the energy-momentum of the gravitational perturbation field is much smaller than the curvature of the background spacetime, the back-reaction on the background metric can be neglected. The perturbation equation is derived from the linearized field equations, with the background metric function $f(r)$ remaining unchanged. This approximation allows a complex nonlinear problem to be simplified into a linear wave equation. When expressed in terms of the metric function \(f(r)\) and its derivative, it can be written as
\begin{equation}
	V_{\rm grav}(r)=f(r)\left[\frac{l(l+1)}{r^2}-\frac{f'(r)}{r}+\frac{2(f(r)-1)}{r^2}\right],
\end{equation}
and when $(1+\alpha)(\alpha M^2+Q^2 )=0$, implying \((1+\alpha)M=M\), the above Eq.~\eqref{eff} reduces to the Regge-Wheeler potential for the Schwarzschild case:
\begin{equation}
	V_{\rm grav}^{\rm Schw}=f(r)\left[\frac{l(l+1)}{r^2}-\frac{6M}{r^3}\right].
\end{equation}

\subsection{Summary}
For RN-MOG black holes, the radial equations governing the perturbations of massless scalar fields, electromagnetic fields, and gravitational fields can all be reduced to the form:
\begin{equation}
	\frac{d^2\psi}{dr_*^2}+[\omega^2-V_{\rm eff}(r)]\psi=0,
\end{equation}
where the effective potentials for massless scalar fields and electromagnetic fields can be expressed in a unified form:
\begin{equation}
	V_{\rm eff} =f(r)\left[\frac{l(l+1)}{r^2}+\frac{(1-s^2)f'(r)}{r}\right],
\end{equation}
with $s=0$ representing the scalar field and $s=1$ representing the electromagnetic field.\\ The effective potential for the perturbations of the gravitational field is:
\begin{equation}
	V_{\rm eff} =f(r)\left[\frac{l(l+1)}{r^2}-\frac{f'(r)}{r}+\frac{s(f(r)-1)}{r^2}\right],
\end{equation}
where, for convenience, we set $s=2$ to represent the gravitational field. Having derived the effective potential for scalar, electromagnetic (vector), and gravitational perturbations, we will proceed to calculate the quasinormal modes in the next section.

The boundary conditions for quasinormal modes require the waves to propagate purely inwards at the event horizon and purely outwards at spatial infinity~\cite{doi:10.1142/S0217751X02011795}:
\begin{equation}
	\psi \sim \begin{cases}
		e^{-i\omega r_*}, & r_* \to -\infty\ (\text{near-horizon}),\\
		e^{+i\omega r_*}, & r_* \to +\infty\ (\text{spatial infinity}).
	\end{cases}
	\label{eq:bc}
\end{equation}

\section{Quasinormal Modes}\label{chapter4}
The WKB method is a well-established approach for calculating the quasinormal modes of black holes, it was originally developed by Wentzel, Kramers, and Brillouin. Schutz and Will~\cite{PhysRevD.35.3621} were the first to apply the WKB method to this problem. Since they initially employed the first-order WKB method, its limited accuracy prompted Iyer and Will~\cite{Kim2013} to extend the method to the third order, a formula that offered significantly improved precision and remained in use for many years. However, the third-order approximation still lacked sufficient accuracy in certain parameter regimes. In 2003, Konoplya extended the calculation of correction terms to the sixth order~\cite{PhysRevD.68.024018}, and subsequently, J. Matyjasek and M. Opala extended it to the thirteenth order~\cite{PhysRevD.96.024011}. Later in 2019, Konoplya, Zhidenko, and Zinhailo proposed a comprehensive scheme incorporating the thirteenth-order WKB method and Padé approximants.

In this chapter, the semi-classical WKB method is employed to calculate the quasinormal frequencies of RN-MOG black holes for massless scalar ($s=0$), electromagnetic ($s=1$), and gravitational perturbations ($s=2$)~\cite{MANFREDI2018492}. The effective potentials and radial master equations for these three types of perturbations were derived in the previous chapter. Subsequently, standard WKB method ranging from the 3rd to the 6th order were used to compute the quasinormal frequencies for various parameters. Finally, the Padé approximation was introduced to extend the WKB series to the 13th order for greater precision, and an error analysis was conducted.

\subsection{WKB approximation}
The basic idea of the WKB method is to perform a Taylor expansion of the effective potential around its peak \(r_0\) and match the asymptotic solutions from both sides of the potential barrier. First-order WKB yields a Bohr-Sommerfeld-type quantization condition for the frequency, while higher-order WKB expresses the frequency \(\omega\) as a perturbative series involving the potential function and its derivatives. Letting \(V_0 = V_{\rm eff}(r_0)\) and denoting \(V_0^{(k)}\) as the \(k\)-th derivative of the potential at the peak, the \(m\)-th order WKB method can be written in a unified form \cite{PhysRevD.96.024011,PhysRevD.68.024018,1985ApJ...291L..33S,PhysRevD.37.3378}:
\begin{equation}
	\frac{i(\omega^2 - V_0)}{\sqrt{-2V_0^{(2)}}} - \sum_{k=2}^{m} \Lambda_k = n + \frac12, \qquad n=0,1,2,\dots
	\label{eq:WKB}
\end{equation}
Here, \(\Lambda_k\) is a polynomial composed of \(V_0^{(j)}\) (where \(j \le 2k\)) and depends on the overtone number \(n\). For \(m=3\) (third-order WKB), the explicit expressions for \(\Lambda_2\) and \(\Lambda_3\) can be found in Ref.~\cite{Bolokhov2026}. Correction terms \(\Lambda_4\), \(\Lambda_5\), and \(\Lambda_6\) (fourth-order to sixth-order) were provided by Konoplya in Ref.~\cite{PhysRevD.68.024018}. In 2017, Matyjasek and Opala extended the method to the 13th-order correction term. For \(\Lambda_6\) through \(\Lambda_{13}\), please refer to~\cite{PhysRevD.96.024011}. Although extensions beyond the 13th order have been developed, the accuracy of the WKB method gradually decreases as the order increases beyond the 6th order, and the method may fail under certain parameter values. In practical calculations, an appropriate order is selected based on requirements. When calculating the complex frequency \(\omega\), one must numerically determine the peak position \(r_0\) of the potential (satisfying \(V'_0=0\)). 

In this paper, we employ the WKB method to calculate quasinormal frequencies up to the sixth order. Tables \ref{label：WKBS0}, \ref{label：WKBS1}, and \ref{label：WKBS2} present the quasinormal frequencies for massless scalar, electromagnetic, and gravitational perturbations, respectively. Additionally, we perform an error analysis for the quasinormal modes (fundamental mode, $n=0$) associated with massless scalar fields ($s=0$), electromagnetic fields ($s=1$), and gravitational fields ($s=2$), with the parameters fixed at $n=0, Q=0.8, \alpha=0.5,$ and $l=2$, using the formula~\cite{Bolokhov2026}:
\begin{equation}
	\Delta =\frac{\omega_{k+1}-\omega_{k-1}}{2}\times100\%.
\end{equation}
The results of this error analysis are summarized in Table \ref{label:WKBerror}. As shown in the table, the error in quasinormal frequencies decreases as the WKB order increases, indicating an improvement in accuracy. Repeated calculations reveal that the WKB method performs optimally at the sixth order. Beyond this order, performance deteriorates, and the method may fail for certain parameter values. Furthermore, the WKB method is applicable only to low-overtone quasinormal frequencies ($n \leq l$). It ceases to be valid when $n > l$, which constitutes a limitation of the method.

\begingroup\squeezetable
\begin{table*}[htbp]
	\centering
	\setlength{\arrayrulewidth}{0.6pt}
	\caption[]{Quasinormal modes of scalar perturbations for different values of $\alpha$, calculated by the WKB approximation, with fixed parameters $Q=0.8$ and $l=2$.}
	\begin{ruledtabular}
		\begin{tabular}{cccccc}			
			$\alpha$              & $n$ & 3rd                      & 4th                      & 5th                      & 6th                      \\ \hline
			\multirow[t]{2}{*}{0}   & 0 & 0.552666 $-$ 0.098344 i & 0.553060 $-$ 0.098274 i  & 0.553072 $-$ 0.098341 i & 0.553052 $-$ 0.098344 i \\
			& 1 & 0.536415 $-$ 0.299190 i   & 0.537110 $-$ 0.298803 i   & 0.537242 $-$ 0.299040 i   & 0.537193 $-$ 0.299067 i  \\
			\multirow[t]{2}{*}{0.2} & 0 & 0.468359 $-$ 0.081652 i & 0.468692 $-$ 0.081594 i & 0.468703 $-$ 0.081655 i & 0.468685 $-$ 0.081658 i \\
			& 1 & 0.455196 $-$ 0.248243 i  & 0.455846 $-$ 0.247889 i  & 0.455963 $-$ 0.248104 i  & 0.455917 $-$ 0.248129 i  \\
			\multirow[t]{2}{*}{0.3} & 0 & 0.435270 $-$ 0.075219 i  & 0.435580 $-$ 0.075165 i  & 0.435590 $-$ 0.075223 i & 0.435573 $-$ 0.075226 i \\
			& 1 & 0.423246 $-$ 0.228619 i  & 0.423878 $-$ 0.228278 i  & 0.423989 $-$ 0.228483 i  & 0.423944 $-$ 0.228507 i  \\
			\multirow[t]{2}{*}{0.5} & 0 & 0.381498 $-$ 0.064920 i & 0.381772 $-$ 0.064873 i & 0.381781 $-$ 0.064926 i & 0.381766 $-$ 0.064928 i \\
			& 1 & 0.371237 $-$ 0.197221 i  & 0.371837 $-$ 0.196903 i  & 0.371936 $-$ 0.197089 i  & 0.371894 $-$ 0.197111 i  \\ 
		\end{tabular}\label{label：WKBS0}
	\end{ruledtabular}
\end{table*}	
\endgroup

\begingroup\squeezetable
\begin{table*}[htbp]
	\centering
	\setlength{\arrayrulewidth}{0.6pt}
	\caption[]{Quasinormal modes of electromagnetic perturbations for different values of $\alpha$, calculated by the WKB approximation, with fixed parameters $Q=0.8$ and $l=2$. }
	\begin{ruledtabular}	
		\begin{tabular}{cccccc}
			$\alpha$               &$n$ & 3rd                      & 4th                      & 5th                      & 6th                      \\  \hline
			\multirow[t]{2}{*}{0}   & 0 & 0.526032 $-$ 0.096871 i & 0.526473 $-$ 0.096790 i & 0.526487 $-$ 0.096871 i & 0.526460 $-$ 0.096876 i  \\
			& 1 & 0.508795 $-$ 0.295083 i  & 0.509588 $-$ 0.294623 i  & 0.509756 $-$ 0.294914 i  & 0.509681 $-$ 0.294958 i  \\
			\multirow[t]{2}{*}{0.2} & 0 & 0.446199 $-$ 0.080448 i & 0.446572 $-$ 0.080381 i & 0.446585 $-$ 0.080454 i & 0.446559 $-$ 0.080459 i \\
			& 1 & 0.432270 $-$ 0.244876 i   & 0.433011 $-$ 0.244457 i  & 0.433162 $-$ 0.244724 i  & 0.433083 $-$ 0.244768 i  \\
			\multirow[t]{2}{*}{0.3} & 0 & 0.414835 $-$ 0.074114 i & 0.415182 $-$ 0.074052 i & 0.415195 $-$ 0.074123 i & 0.415172 $-$ 0.074127 i \\
			& 1 & 0.402127 $-$ 0.225526 i  & 0.402848 $-$ 0.225123 i  & 0.402991 $-$ 0.225379 i  & 0.402925 $-$ 0.225416 i  \\
			\multirow[t]{2}{*}{0.5} & 0 & 0.363826 $-$ 0.063970 i & 0.364133 $-$ 0.063917 i & 0.364144 $-$ 0.063982 i & 0.364124 $-$ 0.063985 i  \\
			& 1 & 0.353008 $-$ 0.194556 i  & 0.353690 $-$ 0.194181 i  & 0.353819 $-$ 0.194416 i  & 0.353766 $-$ 0.194445 i   \\		
		\end{tabular}\label{label：WKBS1}	
	\end{ruledtabular}
\end{table*}	
\endgroup

\begingroup\squeezetable
\begin{table*}[htbp]
	\centering
	\setlength{\arrayrulewidth}{0.6pt}
	\caption[]{Quasinormal modes of gravitational perturbations for different values of $\alpha$, calculated by the WKB approximation, with fixed parameters $Q=0.8$ and $l=2$.}
	\begin{ruledtabular}
		\begin{tabular}{cccccc}	
			$\alpha$               & $n$ & 3rd                      & 4th                      & 5th                      & 6th                      \\ \hline
			\multirow[t]{2}{*}{0}   & 0 & 0.429058 $-$ 0.090804 i & 0.429724 $-$ 0.090664 i & 0.429753 $-$ 0.090802 i & 0.429718 $-$ 0.090809 i \\
			& 1 & 0.406948 $-$ 0.278283 i    & 0.408373 $-$ 0.277312 i    & 0.408794 $-$ 0.277932 i    & 0.408629 $-$ 0.278044 i    \\
			\multirow[t]{2}{*}{0.2} & 0 & 0.363898 $-$ 0.075404 i    & 0.364490 $-$ 0.075281 i    & 0.364522 $-$ 0.075436 i    & 0.364463 $-$ 0.075449 i \\
			& 1 & 0.346045 $-$ 0.230887 i    & 0.347431 $-$ 0.229965 i    & 0.347882 $-$ 0.230645 i    & 0.347590 $-$ 0.230839 i     \\
			\multirow[t]{2}{*}{0.3} & 0 & 0.338301 $-$ 0.069465 i & 0.338863 $-$ 0.069350 i & 0.338895 $-$ 0.069508 i & 0.338829 $-$ 0.069521 i \\
			& 1 & 0.322023 $-$ 0.212625 i    & 0.323384 $-$ 0.211730 i     & 0.323835 $-$ 0.212417 i    & 0.323510 $-$ 0.212631 i     \\
			\multirow[t]{2}{*}{0.5} & 0 & 0.296676 $-$ 0.059956 i    & 0.297185 $-$ 0.059853 i    & 0.297217 $-$ 0.060010 i    & 0.297140 $-$ 0.060026 i    \\
			& 1 & 0.282837 $-$ 0.183400 i      & 0.284143 $-$ 0.182557 i    & 0.284581 $-$ 0.183238 i    & 0.284192 $-$ 0.183489 i    \\ 
		\end{tabular}\label{label：WKBS2}	
	\end{ruledtabular}
\end{table*}	
\endgroup

\begingroup\squeezetable
\begin{table*}[htbp]
	\centering
	\setlength{\arrayrulewidth}{0.6pt}
	\caption[]{Errors of quasinormal modes for scalar ($s=0$), electromagnetic ($s=1$), and gravitational ($s=2$) perturbations, calculated by the WKB approximation, with fixed parameters $n=0$, $Q=0.8$, $\alpha=0.5$, and $l=2$.}
	\begin{ruledtabular}	
		\begin{tabular}{cccc}		
			s                  & WKB order & QNM frequency            & Error estimation($\Delta $) \\ \hline
			\multirow[t]{4}{*}{0} & 3         & 0.381498 $-$ 0.064920 i   & $1.01189\times10^{-3}$ \\
			& 4         & 0.381772 $-$ 0.064873 i   & $1.41273\times10^{-4}$ \\
			& 5         & 0.381781 $-$ 0.064926 i   & $2.76898\times10^{-5}$ \\
			& 6         & 0.381766 $-$ 0.064928 i   & $8.75206\times10^{-6}$ \\
			\multirow[t]{4}{*}{1} & 3         & 0.363826 $-$ 0.063970 i   & $1.07718\times10^{-3}$ \\
			& 4         & 0.364133 $-$ 0.063917 i   & $1.59096\times10^{-4}$ \\
			& 5         & 0.364144 $-$ 0.063982 i   & $3.44937\times10^{-5}$ \\
			& 6         & 0.364124 $-$ 0.063985 i   & $3.85758\times10^{-5}$ \\
			\multirow[t]{4}{*}{2} & 3         & 0.296676 $-$ 0.059956 i   & $1.41937\times10^{-3}$ \\
			& 4         & 0.297185 $-$ 0.059853 i   & $2.71571\times10^{-4}$ \\
			& 5         & 0.297217 $-$ 0.060010 i   & $8.90602\times10^{-5}$ \\
			& 6         & 0.297140 $-$ 0.060026 i   & $6.58492\times10^{-5}$ \\ 
		\end{tabular}\label{label:WKBerror}	
	\end{ruledtabular}
\end{table*}	
\endgroup

\begin{figure*}[htbp]
	\centering
	\begin{minipage}{0.48\textwidth}
		\centering
		\includegraphics[width=\linewidth]{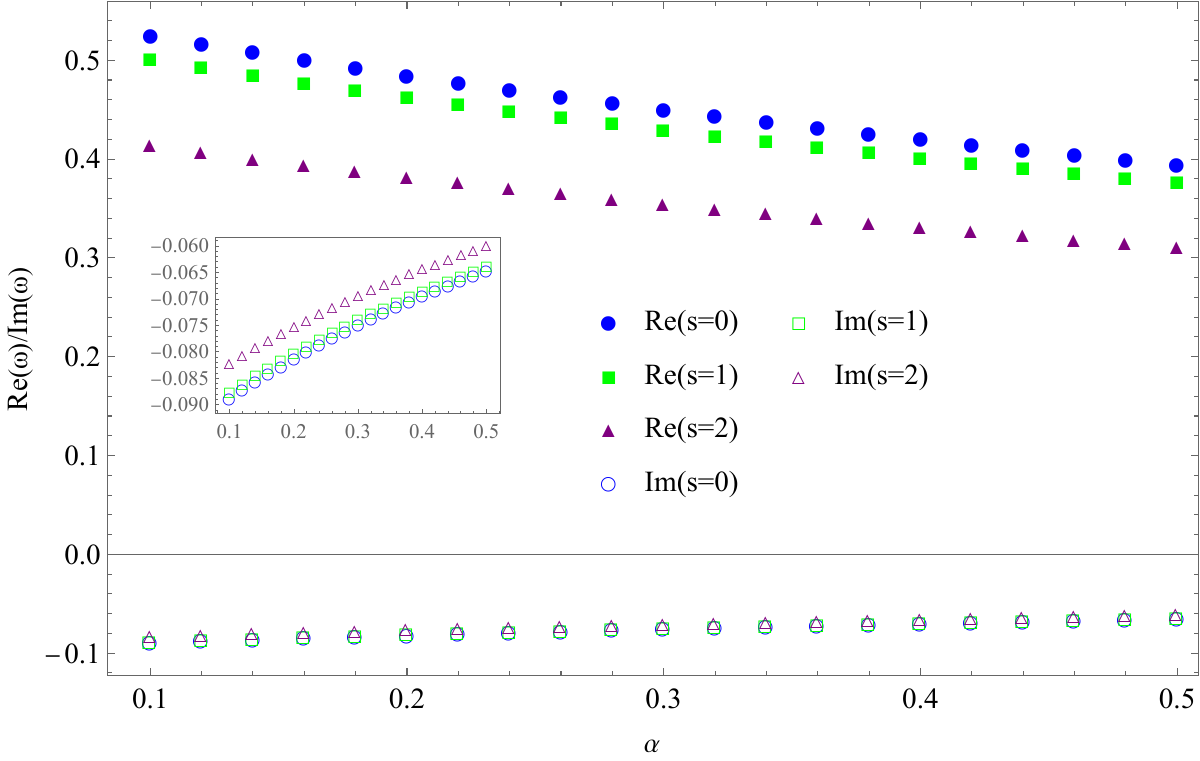}
	\end{minipage}
	\hfill
	\begin{minipage}{0.48\textwidth}
		\centering
		\includegraphics[width=\linewidth]{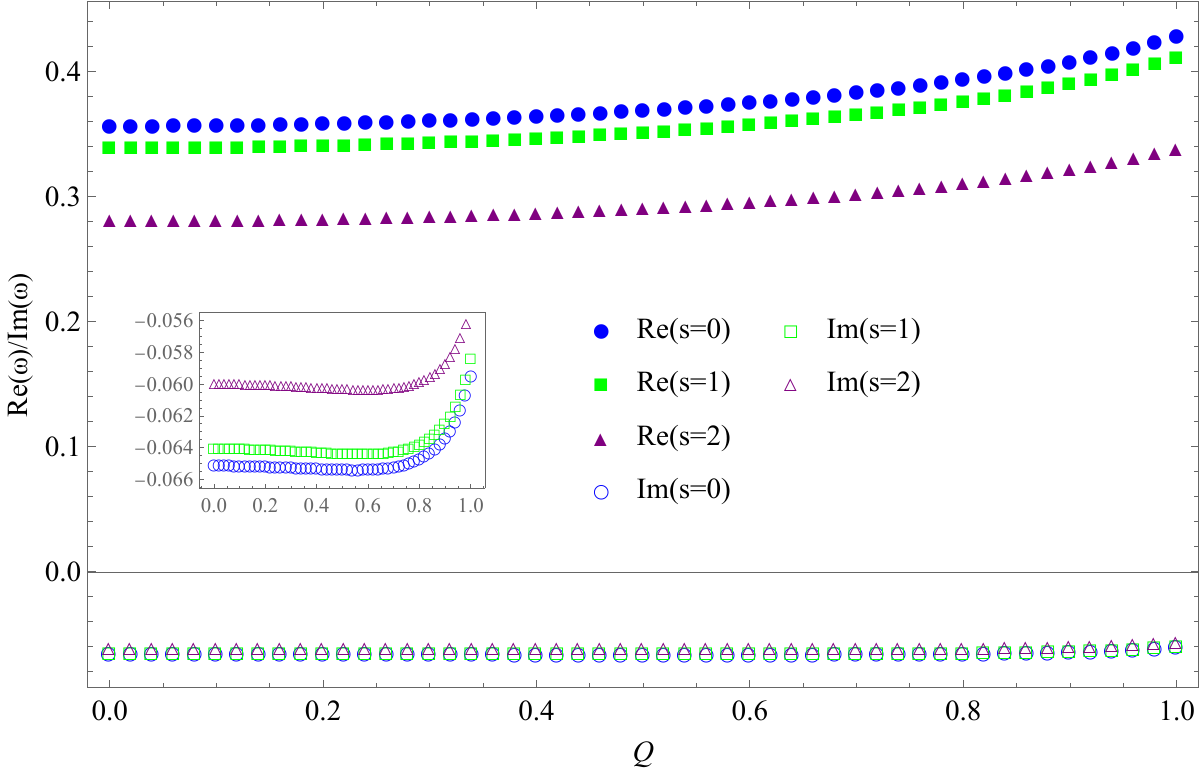}
	\end{minipage}	
	\caption{Trends of the quasinormal frequencies (fundamental mode, $n=0$) for RN-MOG black holes under perturbations by massless scalar fields ($s=0$), electromagnetic fields ($s=1$), and gravitational fields ($s=2$) as functions of the MOG parameter $\alpha$ (left panel) and electric charge $Q$ (right panel). Left panel: $Q=0.8, l=2$. right panel: $\alpha=0.5, l=2$.}
	\label{f6}
\end{figure*}

By fixing specific parameters, we employed the 6th-order WKB method to plot the trends of the quasinormal modes (fundamental frequency, $n=0$) for RN-MOG black holes under three types of perturbations as functions of the modified gravity parameter $\alpha$ and the charge $Q$, as shown in Fig.~\ref{f6}. In the left panel, the charge is fixed at $Q=0.8$ and the angular quantum number at $l=2$. In the right panel, the modified gravity parameter is fixed at $\alpha=0.5$ and the angular quantum number at $l=2$. From the left panel of Fig.~\ref{f6}, we observe that as the modified gravity parameter $\alpha$ increases, the real part of the quasinormal frequency decreases while the imaginary part increases. As shown in the right panel of Fig.~\ref{f6}, the real part of the quasinormal frequency increases with the charge $Q$, whereas the imaginary part decreases. Nevertheless, as $Q$ approaches its critical value, the imaginary part exhibits an increasing trend, which is consistent with the results in Ref.~\cite{PhysRevD.96.024011}. All three types of perturbations follow these patterns.

Let the quasinormal frequency be $\omega = \omega_R + i\omega_I$ (with $\omega_I < 0$). The real part $\omega_R$ describes the oscillation frequency, while the imaginary part $\omega_I$ describes the decay~\cite{Kokkotas1999}. The magnitude of the imaginary part $|\omega_I|$, indicates the rate of decay: a larger $|\omega_I|$ corresponds to faster decay, whereas a smaller $|\omega_I|$ corresponds to slower decay. In the context of quasinormal modes, ``damping'' refers to the phenomenon where the amplitude of gravitational waves (or perturbations in scalar or electromagnetic fields) emitted after a black hole is perturbed decays exponentially over time. $|\omega_I|$ represents the damping rate. From this, we conclude that as the modified gravity parameter $\alpha$ increases, the imaginary part of the quasinormal frequency increases while its magnitude decreases, indicating weaker damping and slower decay of the perturbations. As the charge $Q$ increases, the imaginary part decreases while its magnitude increases, indicating stronger damping and faster decay of the perturbations.

\subsection{Padé approximation}
Since WKB series are usually asymptotic rather than convergent, increasing the order does not guarantee a monotonic improvement in accuracy. To improve convergence and accuracy, we adopt the Padé approximation~\cite{PhysRevD.96.024011} proposed by Matyjasek and Opala. Treat the frequency square vector $P_k(\epsilon)$ to be found as a polynomial with respect to the formal parameter \(\epsilon\):
\begin{equation} 
	P_k(\epsilon)=\omega^2 = V_0 -i \sqrt{-2V_0^{(2)}}\left[ \sum_{i=2}^{k} \Lambda_i \,\epsilon^i + \left(n+\frac12\right) \right]. 
	\label{eq:WKBpk}
\end{equation}
Write $P_k(\epsilon)$ in coefficient form:
\begin{equation} 
	P_k(\epsilon)=c_0+c_1\epsilon+c_2\epsilon^2+\cdots+c_k\epsilon^k, 
	\label{eq:P}
\end{equation}
Comparing Eqs. \eqref{eq:WKBpk} and \eqref{eq:P} gives
\begin{equation}
	\begin{split}
		c_0 &= V_0 - i\sqrt{-2V_0^{(2)}}\Bigl(n+\frac12\Bigr),\\
		c_1 &= 0,\\
		c_i &= -i\sqrt{-2V_0^{(2)}}\,\Lambda_i \quad (i\ge 2).
	\end{split}
	\label{eq:coeffs}
\end{equation}
The $[n/m]$~Pad\'e of the polynomial $P_k(\epsilon)$ is defined as a rational function~\cite{Konoplya_2019}:
\begin{equation} 
	P_{n/m}(\epsilon)=\frac{Q_m(\epsilon)}{R_n(\epsilon)}, 
	\label{eq:pade_def}
\end{equation}
Among them, $Q_m(\epsilon)=\sum_{j=0}^m a_j\epsilon^j$, $R_n(\epsilon)=1+\sum_{j=1}^n b_j\epsilon^j$, and they satisfy
\begin{equation} 
	P_k(\epsilon)-\frac{Q_m(\epsilon)}{R_n(\epsilon)}=O(\epsilon^{k+1}),\qquad k=m+n. 
	\label{eq:order}
\end{equation}
The condition (\ref{eq:order}) is equivalent to requiring the numerator polynomial to be zero:
\begin{equation} 
	R_n(\epsilon)P_k(\epsilon)-Q_m(\epsilon)=O(\epsilon^{k+1}). 
	\label{eq:condition}
\end{equation}
Expand the left side of Eq.~\eqref{eq:condition}:
\begin{equation} 
	\biggl(1+\sum_{j=1}^n b_j\epsilon^j\biggr)\biggl(\sum_{i=0}^k c_i\epsilon^i\biggr)-\sum_{j=0}^m a_j\epsilon^j.
\end{equation}
Let the coefficient of the $\epsilon^\ell$ term be zero ($\ell=0,1,\dots,k$) to obtain a system of linear equations
\begin{equation} 
	c_\ell+\sum_{j=1}^{\min(\ell,n)} b_j\,c_{\ell-j}-a_\ell=0, 
	\qquad\ell=0,1,\dots,k, 
	\label{eq:sys}
\end{equation}
It is stipulated that if $\ell>m$, then $a_\ell=0$, if $\ell-j<0$, then $c_{\ell-j}=0$.
To find the coefficients $b_j$, take $\ell=m+1,\dots,m+n$, so that $a_\ell=0$, and get the equation containing only $b_j$
\begin{equation} 
	c_\ell+\sum_{j=1}^{n} b_j\,c_{\ell-j}=0,\qquad \ell=m+1,\dots,m+n. 
	\label{eq:b_eq}
\end{equation}
Cast into matrix form:
\begin{equation}
	\begin{pmatrix}
		c_{m}      & c_{m-1}    & \cdots & c_{m-n+1} \\[2pt]
		c_{m+1}    & c_{m}      & \cdots & c_{m-n+2} \\[2pt]
		\vdots     & \vdots     & \ddots & \vdots     \\[2pt]
		c_{m+n-1}  & c_{m+n-2}  & \cdots & c_{m}
	\end{pmatrix}
	\begin{pmatrix}
		b_1 \\ b_2 \\ \vdots \\ b_n
	\end{pmatrix}
	= -
	\begin{pmatrix}
		c_{m+1} \\ c_{m+2} \\ \vdots \\ c_{m+n}
	\end{pmatrix}.
	\label{eq:Pade_matrix}
\end{equation}
To find the coefficients $a_j$, for $\ell=0,\dots,m$, they are given explicitly by Eq.~\eqref{eq:sys}
\begin{equation} 
	a_\ell=c_\ell+\sum_{j=1}^{\min(\ell,n)} b_j\,c_{\ell-j}. 
	\label{eq:a_eq}
\end{equation}
After solving for $a_j,b_j$,  the Padé approximant is
\begin{equation} 
	P_{n/m}(\epsilon)=\frac{a_0+a_1\epsilon+\cdots+a_m\epsilon^m} 
	{1+b_1\epsilon+\cdots+b_n\epsilon^n}. 
	\label{eq:pade_explicit}
\end{equation}
Finally, set $\epsilon=1$ to obtain the squared frequency after resuming
\begin{equation} 
	\omega^2\approx P_{n/m}(1).
\end{equation}
In actual calculations, for $k=13$ order WKB, the near-diagonal order $(n,m)=(6,7)$ or $(7,6)$ is often taken, and the deviation of the frequencies obtained from different combinations can be used as an error estimate.

Subsequently, the Padé rational approximation $P_{n/m}(\epsilon)$ is constructed, and its value at $\epsilon=1$ is taken as the new approximation for the squared frequency. The order $(n,m)$ of the Padé approximation satisfies $m+n=k$, with $m=n$ typically chosen. Through this technique, the originally divergent WKB series can be effectively resummed, enabling the use of the otherwise unstable 13th-order WKB method to obtain reliable results \cite{PhysRevD.96.024011,Konoplya_2019}. The Padé average is obtained by averaging all $P_{n/m}$ values calculated at the same order, yielding the Padé average QNM ($\bar{\omega}$).

Here, we calculate frequencies up to the 13th order using the Padé averaging scheme and perform an error analysis of the quasinormal frequencies obtained via Padé averaging at different orders. Here, we define the error in the Padé averaging as the standard deviation of the quasinormal frequencies obtained at the same order:
\begin{equation}
	\delta=\sqrt{\frac{1}{k-1} \sum_{i=1}^{k} (\left| \omega_i - \bar{\omega} \right|)^2}.
\end{equation}

Next, we calculate the quasinormal frequencies from the 1st to the 13th order using the WKB-Padé approximation, determine the average quasinormal frequency for each order via Padé averaging, and calculate the associated errors. The quasinormal modes for massless scalar, electromagnetic, and gravitational perturbations are presented in Tables \ref{padeWKBs0}, \ref{padeWKBs1}, and \ref{padeWKBs2}, respectively.

\begingroup\squeezetable
\begin{table}[htbp]
	\centering
	\setlength{\arrayrulewidth}{0.6pt}
	\caption[]{Quasinormal modes of scalar perturbations for different orders, calculated by the Pad\'e average, with fixed parameters $l=2$, $n=0$, $Q=0.8$, and $\alpha=0.5$.}
	\begin{ruledtabular}	
		\begin{tabular}{ccc}		
			Order & Padé average QNM       & Error estimation($\delta$)  \\ \hline
			1     & 0.384458 $-$ 0.0630099 i & $1.04647\times10^{-2}$               \\
			2     & 0.381709 $-$ 0.0647866 i & $5.76778\times10^{-4}$               \\
			3     & 0.381722 $-$ 0.0648968 i & $1.34162\times10^{-5}$               \\
			4     & 0.381770 $-$ 0.0649168 i  & $6.74195\times10^{-6}$               \\
			5     & 0.381766 $-$ 0.0649244 i & $1.3127\times10^{-7}$                \\
			6     & 0.381766 $-$ 0.0649241 i & $1.44997\times10^{-7}$               \\
			7     & 0.381767 $-$ 0.0649237 i & $1.43636\times10^{-6}$               \\
			8     & 0.381766 $-$ 0.0649241 i & $1.50653\times10^{-7}$               \\
			9     & 0.381769 $-$ 0.0649240 i  & $3.77391\times10^{-6}$               \\
			10    & 0.381766 $-$ 0.0649238 i & $4.49706\times10^{-7}$               \\
			11    & 0.381770 $-$ 0.0649238 i  & $4.56435\times10^{-6}$               \\
			12    & 0.381766 $-$ 0.0649241 i & $3.12467\times10^{-7}$               \\
			13    & 0.381771 $-$ 0.0649242 i & $5.39182\times10^{-6}$               \\ 
		\end{tabular}\label{padeWKBs0}	
	\end{ruledtabular}
\end{table}
\endgroup

\begingroup\squeezetable
\begin{table}[htbp]
	\centering
	\setlength{\arrayrulewidth}{0.6pt}
	\caption[]{Quasinormal modes of electromagnetic perturbations for different orders, calculated by the Pad\'e average, with fixed parameters $l=2$, $n=0$, $Q=0.8$, and $\alpha=0.5$.}
		\begin{ruledtabular}
		\begin{tabular}{ccc}		
			Order & Padé average QNM       & Error estimation($\delta$)  \\ \hline
			1     & 0.366935 $-$ 0.0619889 i & $1.06206\times10^{-2}$               \\
			2     & 0.364064 $-$ 0.0638266 i & $5.99195\times10^{-4}$               \\
			3     & 0.364076 $-$ 0.0639448 i & $1.41345\times10^{-5}$               \\
			4     & 0.364131 $-$ 0.0639684 i & $8.28534\times10^{-6}$               \\
			5     & 0.364125 $-$ 0.0639791 i & $3.64083\times10^{-7}$               \\
			6     & 0.364125 $-$ 0.0639789 i & $9.20314\times10^{-10}$              \\
			7     & 0.364128 $-$ 0.0639783 i & $5.64024\times10^{-6}$               \\
			8     & 0.364124 $-$ 0.0639777 i & $5.9432\times10^{-7}$                \\
			9     & 0.364128 $-$ 0.0639783 i & $5.57249\times10^{-6}$               \\
			10    & 0.364125 $-$ 0.0639785 i & $4.57117\times10^{-7}$               \\
			11    & 0.364131 $-$ 0.0639786 i & $7.25002\times10^{-6}$               \\
			12    & 0.364126 $-$ 0.0639792 i & $4.06203\times10^{-7}$               \\
			13    & 0.364132 $-$ 0.0639791 i & $9.34302\times10^{-6}$               \\ 
		\end{tabular}\label{padeWKBs1}	
	\end{ruledtabular}
\end{table}	
\endgroup

\begingroup\squeezetable
\begin{table}[htbp]
	\centering
	\setlength{\arrayrulewidth}{0.6pt}
	\caption[]{Quasinormal modes of gravitational perturbations for different orders, calculated by the Pad\'e average, with fixed parameters $l=2$, $n=0$, $Q=0.8$, and $\alpha=0.5$.}
	\begin{ruledtabular}
		\begin{tabular}{ccc}		
			Order & Padé average QNM       & Error estimation($\delta$) \\ \hline
			1     & 0.300546 $-$ 0.0576760 i    & $1.12702\times10^{-2}$ \\
			2     & 0.297076 $-$ 0.0597632 i   & $6.96318\times10^{-4}$ \\
			3     & 0.297082 $-$ 0.0599171 i   & $1.74605\times10^{-5}$ \\
			4     & 0.297177 $-$ 0.0599652 i   & $2.02849\times10^{-5}$ \\
			5     & 0.297152 $-$ 0.0599963 i   & $3.86041\times10^{-6}$ \\
			6     & 0.297150 $-$ 0.0599911 i    & $3.64668\times10^{-7}$ \\
			7     & 0.297160 $-$ 0.0599930 i     & $1.36959\times10^{-5}$ \\
			8     & 0.297150 $-$ 0.0599898 i    & $1.61083\times10^{-6}$ \\
			9     & 0.297158 $-$ 0.0599923 i   & $1.15397\times10^{-5}$ \\
			10    & 0.297150 $-$ 0.0599904 i    & $2.10071\times10^{-6}$ \\
			11    & 0.297164 $-$ 0.0599916 i   & $1.63195\times10^{-5}$ \\
			12    & 0.297150 $-$ 0.0599868 i    & $2.6545\times10^{-6}$ \\
			13    & 0.297164 $-$ 0.0599899 i   & $1.66752\times10^{-5}$ \\  
		\end{tabular}\label{padeWKBs2}	
	\end{ruledtabular}
\end{table}	
\endgroup

As shown in Tables \ref{padeWKBs0}-\ref{padeWKBs2}, the quasinormal frequencies calculated using the Padé approximation are superior to those obtained via the WKB approximation. In particular, the frequency error decreases at higher orders, thereby compensating for the limitations of the WKB approximation. Given the low errors associated with the 6th, 8th, 10th, 12th, and 13th-order Padé approximations, we employed these orders to calculate the quasinormal frequencies for higher overtones ($n=0, 1, 2$) and higher angular quantum numbers ($l=2, 3, 4$). The quasinormal modes for massless scalar, electromagnetic, and gravitational perturbations of RN-MOG black holes, calculated using the Padé approximation, are presented in Tables \ref{Pnm0}, \ref{Pnm1}, and \ref{Pnm2}, respectively.

As can be seen from Tables \ref{Pnm0}-\ref{Pnm2}, the magnitude of the imaginary part of the quasinormal frequency increases with the overtone number ($n$), indicating enhanced damping and more rapid decay of perturbations. Regarding the angular quantum number ($l$), it significantly affects the real part but has a relatively minor impact on the imaginary part. This suggests that the angular quantum number primarily influences the oscillation frequency while having a smaller effect on the decay rate.

\begingroup\squeezetable
\begin{table*}[htbp]
	\centering
	\setlength{\arrayrulewidth}{0.6pt}
	\caption[]{Quasinormal modes of scalar perturbations for different overtones, calculated by the Pad\'e approximation, with fixed parameters $Q=0.8$ and $\alpha=0.5$.}
	\begin{ruledtabular}	
		\begin{tabular}{ccccc}	
			$l$                 & Padé approximation & $n=0$                      &$n=1$                 &$n=2$                   \\ \hline
			\multirow[t]{5}{*}{2} & $P_{3/3}$   & 0.381766 $-$ 0.0649241 i & 0.371892 $-$ 0.197107 i & 0.354741 $-$ 0.335425 i \\
			& $P_{4/4}$   & 0.381766 $-$ 0.0649241 i & 0.371892 $-$ 0.197107 i & 0.354741 $-$ 0.335424 i \\
			& $P_{5/5}$   & 0.381766 $-$ 0.0649238 i & 0.371886 $-$ 0.197108 i & 0.354730 $-$ 0.335450 i   \\
			& $P_{6/6}$   & 0.381766 $-$ 0.0649242 i & 0.371892 $-$ 0.197107 i & 0.354741 $-$ 0.335426 i \\
			& $P_{6/7}$   & 0.381766 $-$ 0.0649244 i & 0.371876 $-$ 0.197110 i & 0.354042 $-$ 0.336173 i \\
			\multirow[t]{5}{*}{3} & $P_{3/3}$   & 0.533018 $-$ 0.0648087 i & 0.525719 $-$ 0.195649 i & 0.512122 $-$ 0.329934 i \\
			& $P_{4/4}$   & 0.533018 $-$ 0.0648088 i & 0.525723 $-$ 0.195648 i & 0.512139 $-$ 0.329912 i \\
			& $P_{5/5}$   & 0.533018 $-$ 0.0648082 i & 0.525706 $-$ 0.195647 i & 0.512038 $-$ 0.329994 i \\
			& $P_{6/6}$   & 0.533018 $-$ 0.0648084 i & 0.525712 $-$ 0.195649 i & 0.512091 $-$ 0.329964 i \\
			& $P_{6/7}$   & 0.533018 $-$ 0.0648088 i & 0.525717 $-$ 0.195649 i & 0.511942 $-$ 0.330006 i \\
			\multirow[t]{5}{*}{4} & $P_{3/3}$   & 0.684543 $-$ 0.0647610 i & 0.678786 $-$ 0.195031 i & 0.667753 $-$ 0.327461 i \\
			& $P_{4/4}$   & 0.684543 $-$ 0.0647611 i & 0.678789 $-$ 0.195030 i & 0.667765 $-$ 0.327452 i \\
			& $P_{5/5}$   & 0.684544 $-$ 0.0647611 i & 0.678790 $-$ 0.195030 i & 0.667776 $-$ 0.327444 i \\
			& $P_{6/6}$   & 0.684543 $-$ 0.0647610 i & 0.678787 $-$ 0.195030 i & 0.667757 $-$ 0.327458 i \\
			& $P_{6/7}$   & 0.684543 $-$ 0.0647610 i & 0.678785 $-$ 0.195031 i & 0.667758 $-$ 0.327502 i \\ 
		\end{tabular}\label{Pnm0}	
	\end{ruledtabular}
\end{table*}	
\endgroup

\begingroup\squeezetable
\begin{table*}[htbp]
	\centering
	\setlength{\arrayrulewidth}{0.6pt}
	\caption[]{Quasinormal modes of electromagnetic perturbations for different overtones, calculated by the Pad\'e approximation, with fixed parameters $Q=0.8$ and $\alpha=0.5$. }
	\begin{ruledtabular}
		\begin{tabular}{ccccc}	
			$l$                  & Padé approximation &$n=0$                     &$n=1$                 & $n=2$                 \\ \hline
			\multirow[t]{5}{*}{2} & $P_{3/3}$ & 0.364125 $-$ 0.0639789 i & 0.353764 $-$ 0.194439 i & 0.335871 $-$ 0.331425 i \\
			& $P_{4/4}$ & 0.364124 $-$ 0.0639776 i & 0.353743 $-$ 0.194443 i & 0.335828 $-$ 0.331522 i \\
			& $P_{5/5}$ & 0.364125 $-$ 0.0639786 i & 0.353757 $-$ 0.194441 i & 0.335858 $-$ 0.331457 i \\
			& $P_{6/6}$ & 0.364125 $-$ 0.0639794 i & 0.353773 $-$ 0.194436 i & 0.335890 $-$ 0.331371 i \\
			& $P_{6/7}$ & 0.364125 $-$ 0.0639790 i & 0.353739 $-$ 0.194446 i & 0.335230 $-$ 0.332662 i \\
			\multirow[t]{5}{*}{3} & $P_{3/3}$ & 0.520478 $-$ 0.0643263 i & 0.512997 $-$ 0.194244 i & 0.499079 $-$ 0.327720 i \\
			& $P_{4/4}$ & 0.520478 $-$ 0.0643261 i & 0.512993 $-$ 0.194244 i & 0.499065 $-$ 0.327736 i \\
			& $P_{5/5}$ & 0.520478 $-$ 0.0643262 i & 0.512996 $-$ 0.194244 i & 0.499074 $-$ 0.327725 i \\
			& $P_{6/6}$ & 0.520478 $-$ 0.0643263 i & 0.512999 $-$ 0.194243 i & 0.499086 $-$ 0.327711 i \\
			& $P_{6/7}$ & 0.520478 $-$ 0.0643263 i & 0.512992 $-$ 0.194245 i & 0.498953 $-$ 0.327941 i \\
			\multirow[t]{5}{*}{4} & $P_{3/3}$ & 0.674809 $-$ 0.0644690 i & 0.668964 $-$ 0.194170 i & 0.657766 $-$ 0.326080 i \\
			& $P_{4/4}$ & 0.674809 $-$ 0.0644690 i & 0.668964 $-$ 0.194180 i & 0.657766 $-$ 0.326080 i \\
			& $P_{5/5}$ & 0.674809 $-$ 0.0644690 i & 0.668964 $-$ 0.194190 i & 0.657764 $-$ 0.326081 i \\
			& $P_{6/6}$ & 0.674809 $-$ 0.0644690 i & 0.668964 $-$ 0.194200 i & 0.657765 $-$ 0.326081 i \\
			& $P_{6/7}$ & 0.674809 $-$ 0.0644690 i & 0.668964 $-$ 0.194210 i & 0.657708 $-$ 0.326088 i \\ 
		\end{tabular}\label{Pnm1}	
	\end{ruledtabular}
\end{table*}	
\endgroup

\begingroup\squeezetable
\begin{table*}[htbp]
	\centering
	\setlength{\arrayrulewidth}{0.6pt}
	\caption[]{Quasinormal modes of gravitational perturbations for different overtones, calculated by the Pad\'e approximation, with fixed parameters $Q=0.8$ and $\alpha=0.5$. }
	\begin{ruledtabular}
		\begin{tabular}{ccccc}	
			$l$                  & Padé approximation & $n=0$                      &$n=1$                  & $n=2$                   \\ \hline
			\multirow[t]{5}{*}{2} & $P_{3/3}$ & 0.297149 $-$ 0.0599911 i & 0.284145 $-$ 0.183330 i  & 0.262350 $-$ 0.316199 i  \\
			& $P_{4/4}$ & 0.297149 $-$ 0.0599900 i     & 0.284134 $-$ 0.183325 i & 0.262309 $-$ 0.316243 i \\
			& $P_{5/5}$ & 0.297149 $-$ 0.0599905 i   & 0.284135 $-$ 0.183325 i & 0.262309 $-$ 0.316243 i \\
			& $P_{6/6}$ & 0.297148 $-$ 0.0599862 i   & 0.284319 $-$ 0.183327 i & 0.262655 $-$ 0.315710 i  \\
			& $P_{6/7}$ & 0.297151 $-$ 0.0599948 i   & 0.284270 $-$ 0.183344 i  & 0.260640 $-$ 0.316648 i  \\
			\multirow[t]{5}{*}{3} & $P_{3/3}$ & 0.474896 $-$ 0.0625352 i   & 0.466601 $-$ 0.189047 i  & 0.451210 $-$ 0.319632 i  \\
			& $P_{4/4}$ & 0.474896 $-$ 0.0625348 i   & 0.466594 $-$ 0.189047 i & 0.451183 $-$ 0.319659 i \\
			& $P_{5/5}$ & 0.474896 $-$ 0.0625352 i   & 0.466601 $-$ 0.189047 i & 0.451211 $-$ 0.319631 i \\
			& $P_{6/6}$ & 0.474896 $-$ 0.0625350 i    & 0.466598 $-$ 0.189047 i & 0.451194 $-$ 0.319650 i  \\
			& $P_{6/7}$ & 0.474896 $-$ 0.0625353 i   & 0.466598 $-$ 0.189047 i & 0.450928 $-$ 0.319728 i \\
			\multirow[t]{5}{*}{4} & $P_{3/3}$ & 0.640046 $-$ 0.0634516 i   & 0.633830 $-$ 0.191178 i  & 0.621925 $-$ 0.321292 i \\
			& $P_{4/4}$ & 0.640046 $-$ 0.0634516 i   & 0.633830 $-$ 0.191178 i  & 0.621927 $-$ 0.321291 i \\
			& $P_{5/5}$ & 0.640046 $-$ 0.0634516 i   & 0.633830 $-$ 0.191178 i  & 0.621928 $-$ 0.321291 i \\
			& $P_{6/6}$ & 0.640046 $-$ 0.0634516 i   & 0.633830 $-$ 0.191178 i  & 0.621925 $-$ 0.321292 i \\
			& $P_{6/7}$ & 0.640046 $-$ 0.0634517 i   & 0.633829 $-$ 0.191178 i & 0.621855 $-$ 0.321326 i   \\ 
		\end{tabular}\label{Pnm2}	
	\end{ruledtabular}
\end{table*}	
\endgroup

\section{Time-domain evolution}\label{chapter5}
Quasinormal modes analysis in the frequency domain yields the characteristic frequencies and damping rates of black holes. However, a complete understanding of the temporal evolution of perturbations requires directly solving the wave equation in the time domain. Time-domain evolution not only verifies the accuracy of frequency-domain QNM results but also reveals physical phenomena that frequency-domain methods cannot capture, specifically, the initial perturbation phase, the ringdown phase (quasinormal modes phase), and the late-time power-law tail~\cite{cl3k-3xt2,Song_2024}. Particularly within the framework of modified gravity, the influence of MOG parameters on signal duration, decay rates, and tail indices is most intuitively demonstrated through time-domain waveforms. In this section, the finite-difference method is employed to solve the time-domain wave equation~\cite{Lin_2023,doi:10.1142/S0129183111016415,Skvortsova_2026}.

In terms of the tortoise coordinate \(r_*\), the perturbation \(\Psi(t,r_*)\) satisfies the one-dimensional wave equation
\begin{equation}
	\frac{\partial^2\Psi}{\partial t^2} - \frac{\partial^2\Psi}{\partial r_*^2} + V_{\rm eff}(r_*)\Psi = 0,
	\label{eq:wave}
\end{equation}
where the effective potential \(V_{\rm eff}(r_*)\) is determined by the metric and the type of perturbation. For numerical solution, we introduce light-cone coordinates~\cite{LUTFUOGLU2026170360}:
\begin{equation}
	u = t - r_*, \qquad v = t + r_*,
\end{equation}
whereupon the spacetime line element becomes \(ds^2 = -du\,dv + r^2 d\Omega^2\), and Eq.~\eqref{eq:wave} transforms into
\begin{equation}
	\frac{\partial^2\Psi}{\partial u\partial v} + \frac{1}{4} V_{\rm eff}(r_*)\Psi = 0,
	\label{eq:wave_uv}
\end{equation}
with \(r_*\) related to \(u\) and \(v\) via \((v-u)/2\). This form eliminates mixed partial derivatives, facilitating the construction of an explicit recurrence scheme.
We select a uniform step size \(\Delta\) in the \(u\)--\(v\) plane and denote the coordinates of discrete points as
\begin{equation}
    u_i = i\Delta,\quad v_j = j\Delta,\qquad i,j = 0,1,2,\dots
\end{equation}
The values at the grid points are denoted as \(\Psi_{i,j} \equiv \Psi(u_i, v_j)\). Using the central difference approximation for the mixed partial derivative:
\begin{equation}
    \frac{\partial^2\Psi}{\partial u\partial v}\bigg|_{i,j} \approx \frac{\Psi_{i,j} - \Psi_{i-1,j} - \Psi_{i,j-1} + \Psi_{i-1,j-1}}{\Delta^2} + O(\Delta^2).
\end{equation}
The potential energy term is taken as the average of four adjacent points to maintain symmetry and stability:
\begin{equation}
	(V\Psi)_{i,j} \approx \frac{1}{4} V_{i-\frac12,j-\frac12} \left(\Psi_{i-1,j} + \Psi_{i,j-1} + \Psi_{i-1,j-1} + \Psi_{i,j}\right),
\end{equation}
where \(V_{i-\frac{1}{2},j-\frac{1}{2}}\) is the value of the potential function at the grid point \((i-\frac{1}{2}, j-\frac{1}{2})\), and the corresponding tortoise coordinate is
\begin{equation}
     r_* = \frac{v_{j-\frac{1}{2}} - u_{i-\frac{1}{2}}}{2} = \frac{(j-i)\Delta}{2}.
\end{equation}
Substituting the difference approximations into Eq.~\eqref{eq:wave_uv} and rearranging yields the recurrence relation
\begin{equation}
	\Psi_{i,j} = \frac{1 - \frac{\Delta^2}{16} V}{1 + \frac{\Delta^2}{16} V}\left(\Psi_{i-1,j} + \Psi_{i,j-1}\right) - \Psi_{i-1,j-1},
	\label{eq:recursion}
\end{equation}
where \(V\) is evaluated at the center point \(((i-\frac{1}{2})\Delta, (j-\frac{1}{2})\Delta)\). This scheme is explicit, allowing for the calculation of new $\Psi_{i,j}$ values by advancing layer by layer in the direction of increasing $(u, v)$.

Time evolution requires specifying initial values \(\Psi(0,r_*)\) and \(\partial_t\Psi(0,r_*)\) at \(t=0\)~\cite{Yang_2024}. In light-cone coordinates, \(t=0\) corresponds to the line \(v = -u\), namely, the diagonal. We choose the initial state to be a purely spatial Gaussian wave packet with a vanishing time derivative~\cite{Macedo_2018,doi:10.1142/S0218271820410060,Corr_a_2024}:
\begin{equation}
   \Psi(0,r_*) = A\exp\left[-\frac{(r_* - a)^2}{2\sigma^2}\right], \qquad \partial_t\Psi(0,r_*) = 0.
\end{equation}
For the numerical implementation on the grid, values matching the initial conditions are assigned to the boundaries at \(u=0\) and \(v=0\). By utilizing symmetry, we set
\begin{equation}
     \Psi(u=0, v) = \Psi(0, r_*),\quad \Psi(u, v=0) = \Psi(0, r_*).
\end{equation}

\begin{figure*}[htbp]
	\centering	
	\begin{minipage}{0.3\textwidth}
		\centering
		\includegraphics[width=\linewidth]{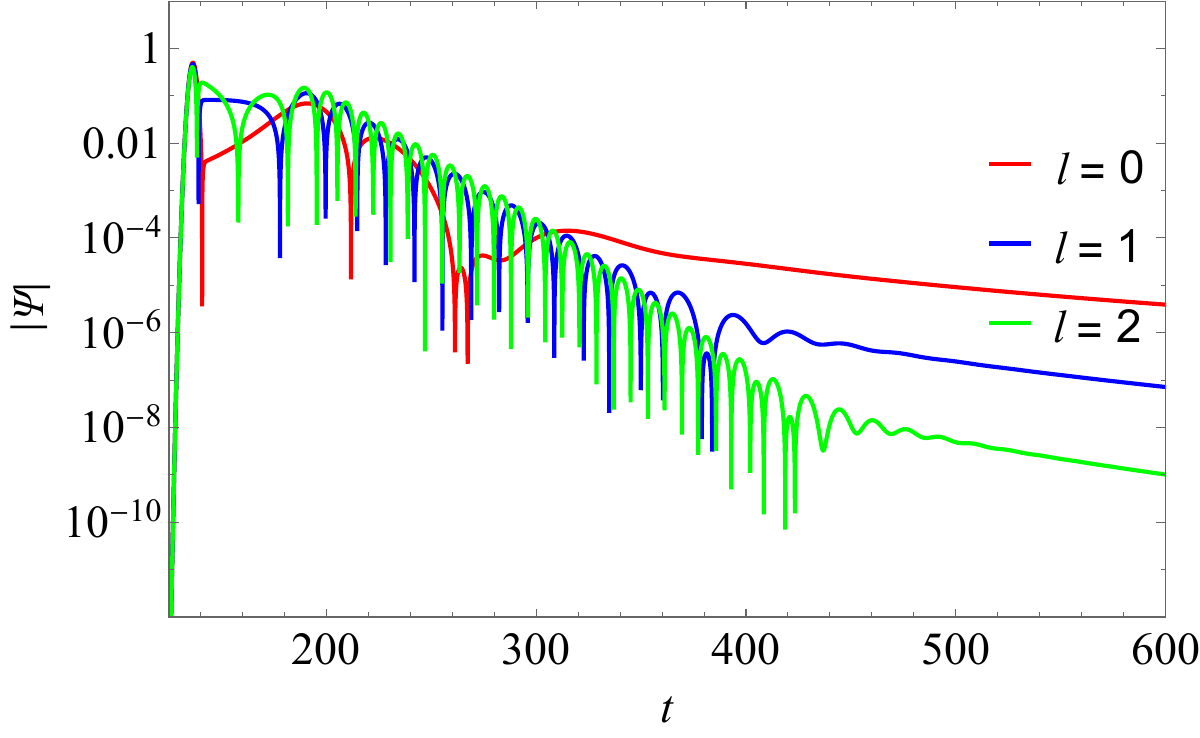}
		\medskip
		Scalar perturbation
	\end{minipage}
	\hfill
	\begin{minipage}{0.3\textwidth}
		\centering
		\includegraphics[width=\linewidth]{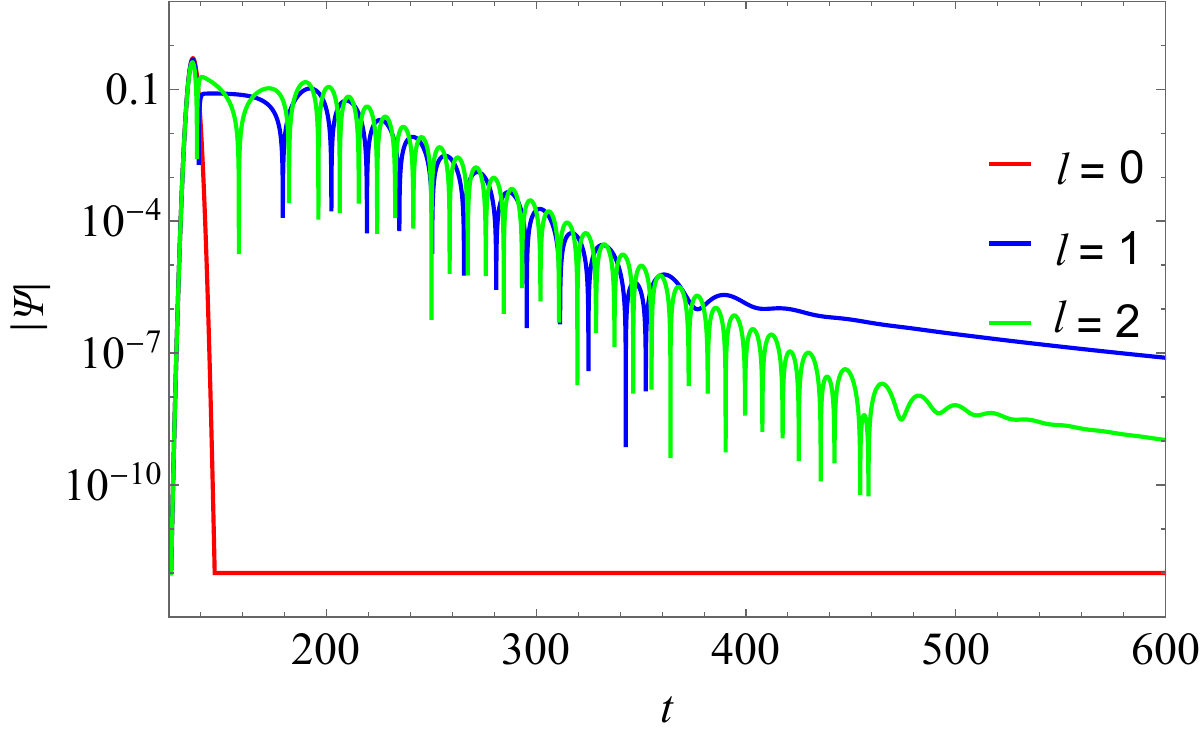}
		\medskip
		Electromagnetic perturbation
	\end{minipage}
	\hfill
	\begin{minipage}{0.3\textwidth}
		\centering
		\includegraphics[width=\linewidth]{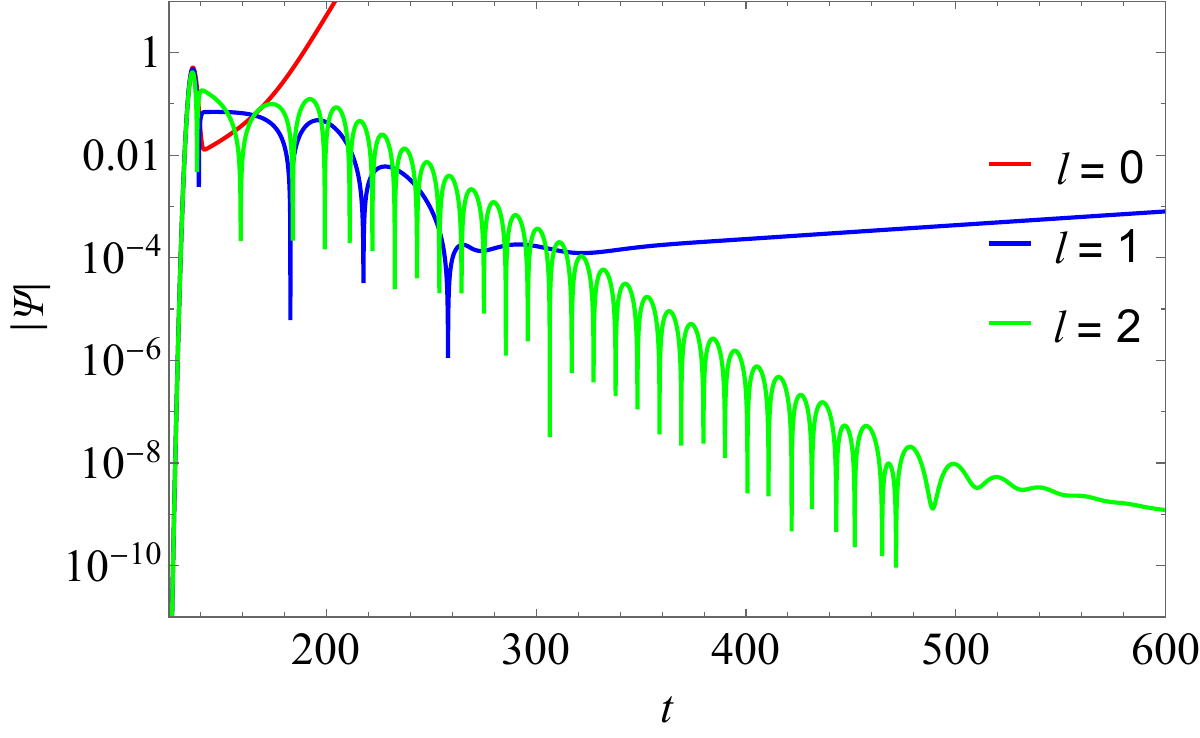}
		\medskip
		Gravitational perturbation
	\end{minipage}
	\caption{Semi-logarithmic plots of the time-domain evolution for three types of perturbations ($l=0, 1, 2$) with $Q=0.8$ and $\alpha=0.5$.}
	\label{f7}
\end{figure*}

\begin{figure*}[htbp]
	\centering	
	\begin{minipage}{0.3\textwidth}
		\centering
		\includegraphics[width=\linewidth]{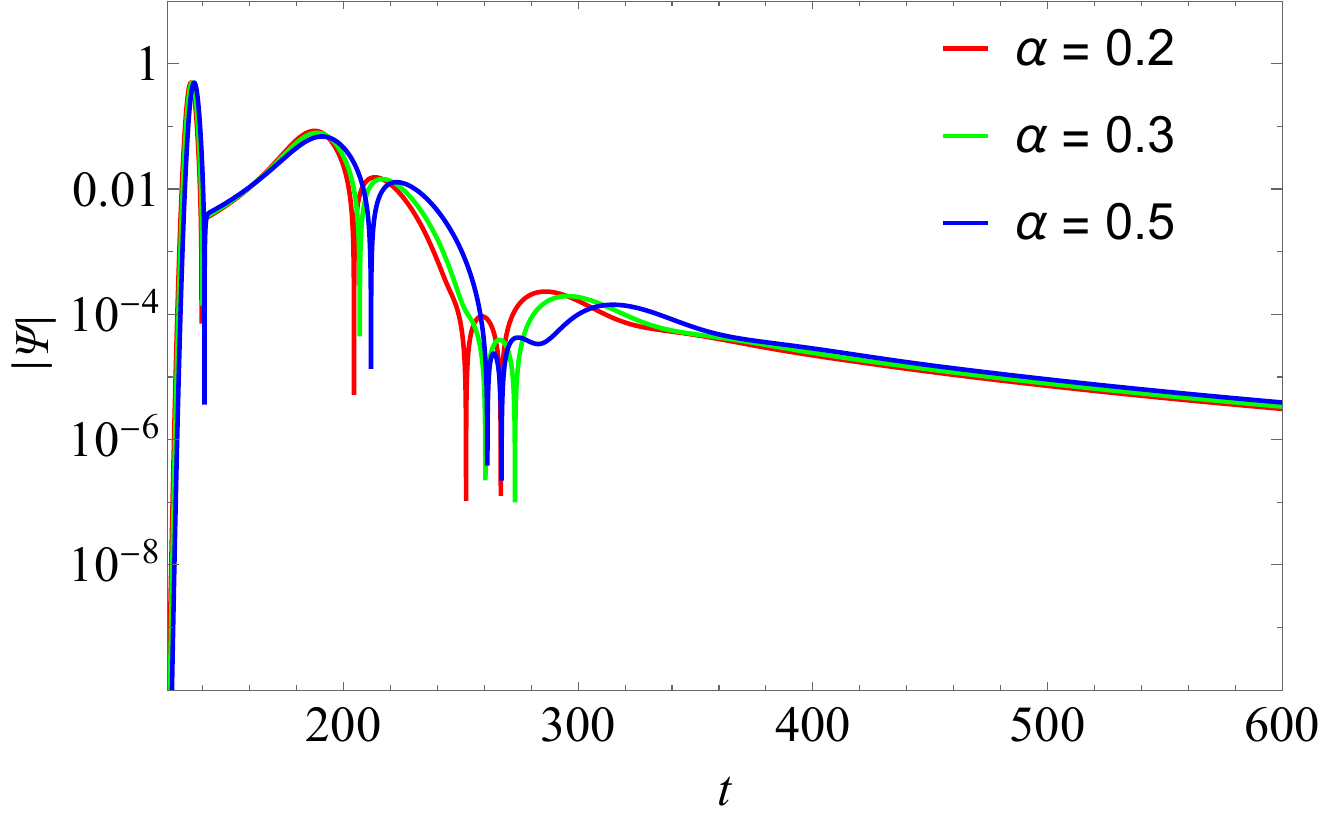}
		\medskip
		$l=0$
	\end{minipage}
	\hfill
	\begin{minipage}{0.3\textwidth}
		\centering
		\includegraphics[width=\linewidth]{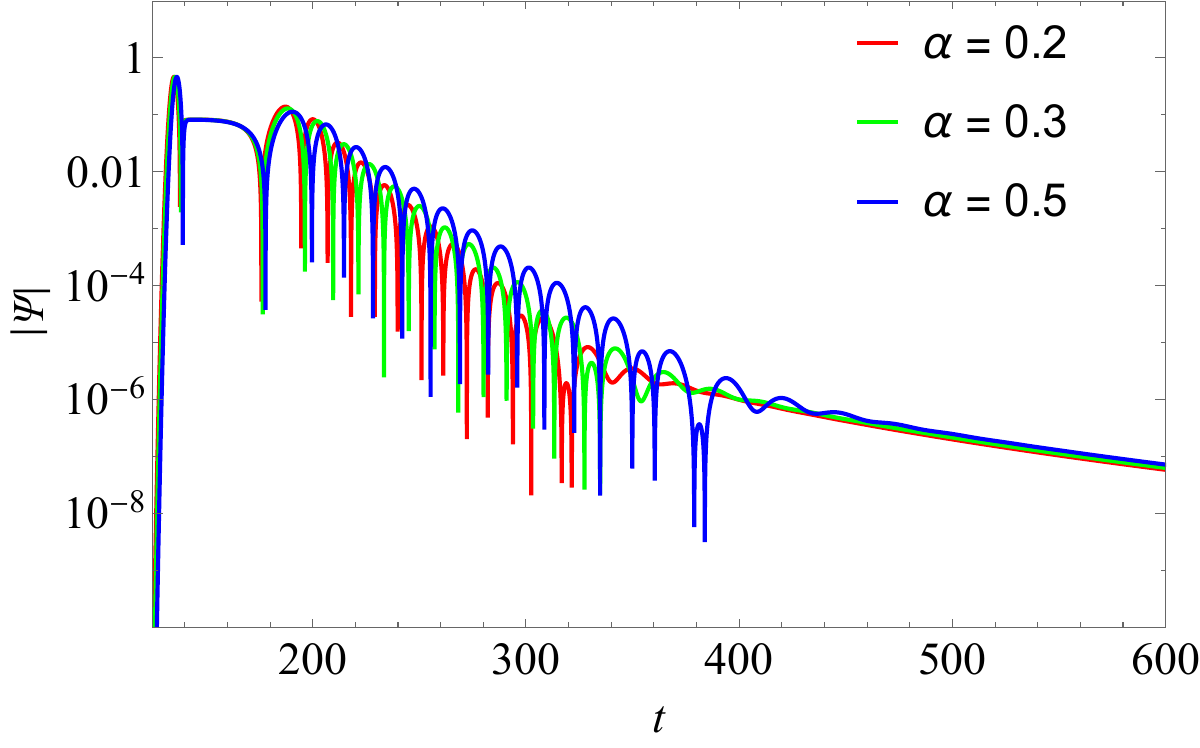}
		\medskip
		$l=1$
	\end{minipage}
	\hfill
	\begin{minipage}{0.3\textwidth}
		\centering
		\includegraphics[width=\linewidth]{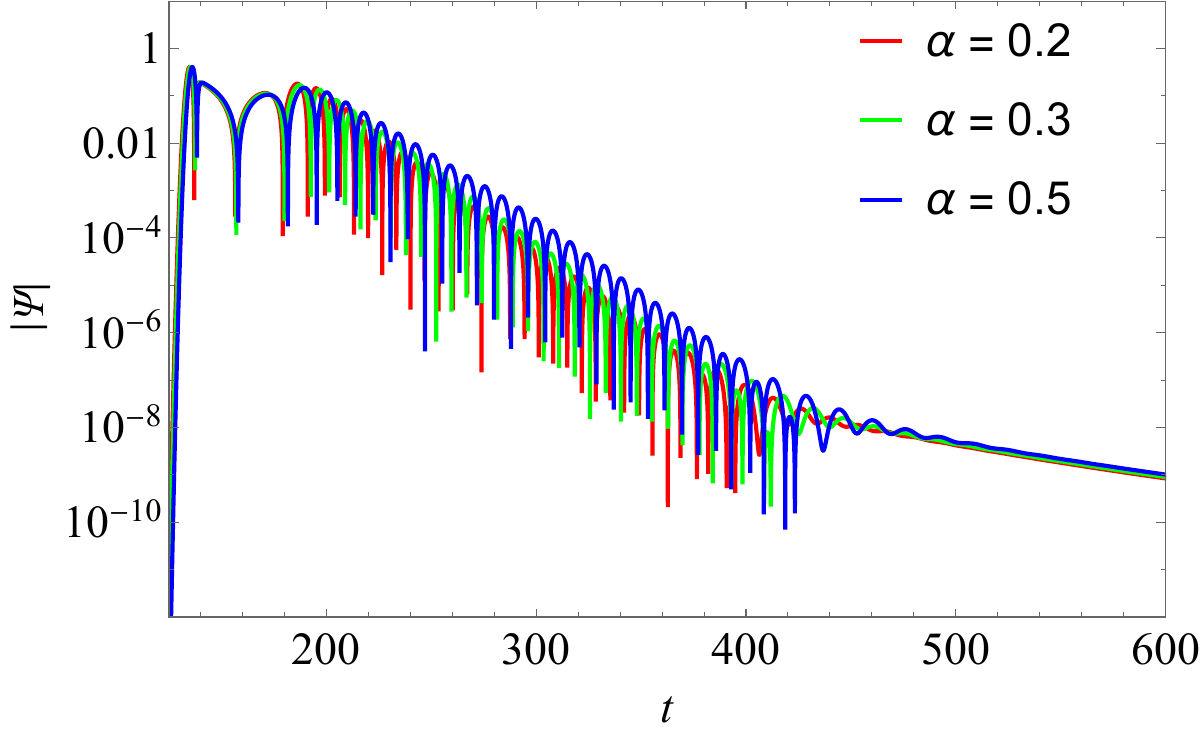}
		\medskip
		$l=2$
	\end{minipage}
	\caption{Semi-logarithmic plot of the time-domain evolution of scalar perturbations ($\alpha=0.2, 0.3, 0.5$) with $Q=0.8$.}
	\label{f8}
\end{figure*}

\begin{figure*}[htbp]
	\centering	
	\begin{minipage}{0.3\textwidth}
		\centering
		\includegraphics[width=\linewidth]{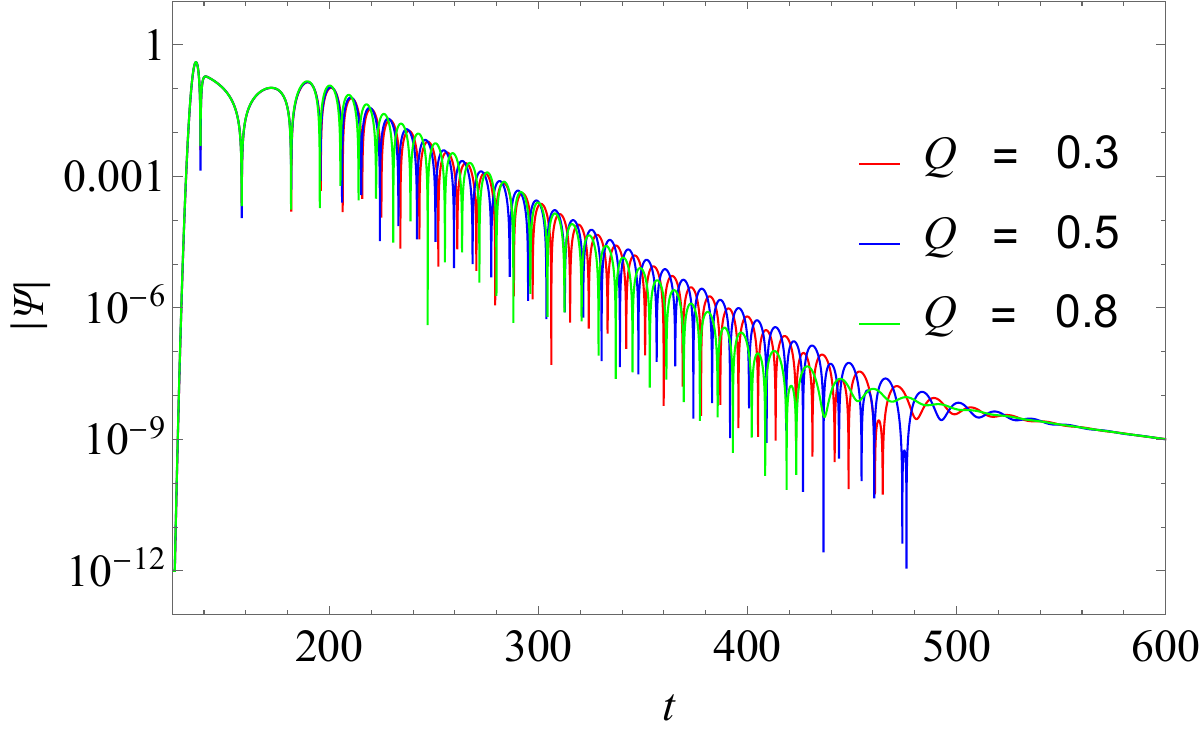}
		\medskip
		Scalar perturbation
	\end{minipage}
	\hfill
	\begin{minipage}{0.3\textwidth}
		\centering
		\includegraphics[width=\linewidth]{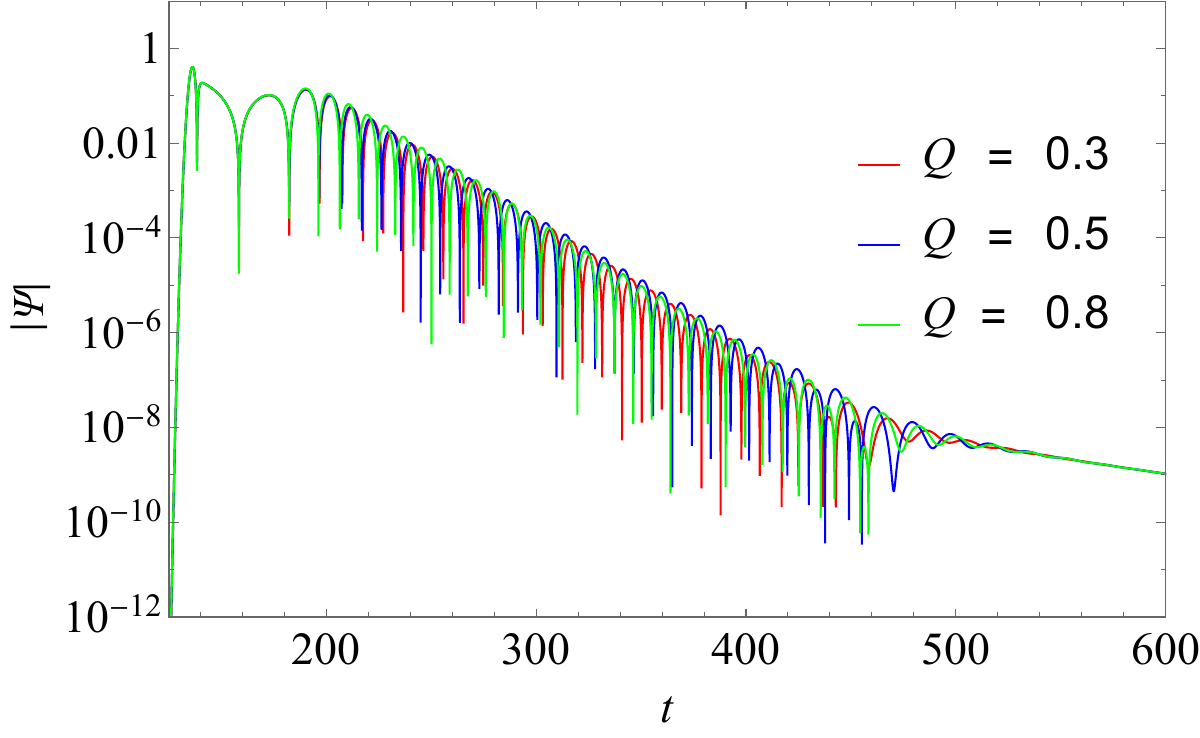}
		\medskip
		Electromagnetic perturbation
	\end{minipage}
	\hfill
	\begin{minipage}{0.3\textwidth}
		\centering
		\includegraphics[width=\linewidth]{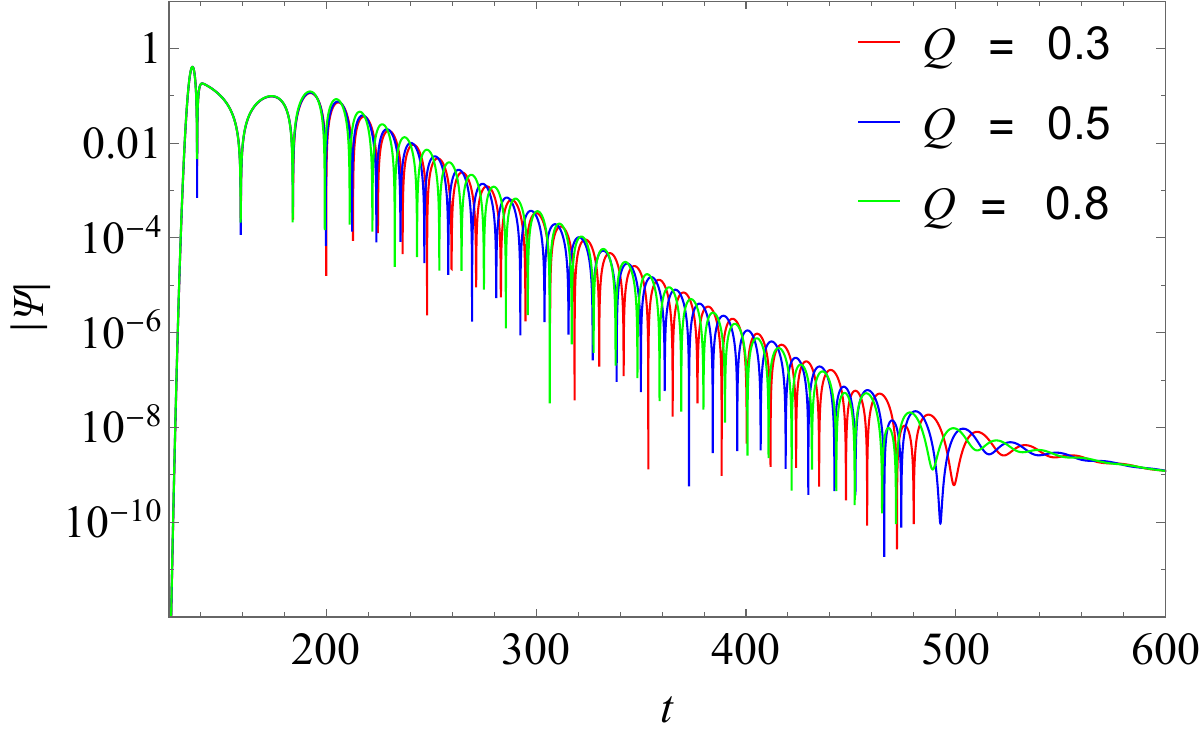}
		\medskip
		Gravitational perturbation
	\end{minipage}
	\caption{Semi-logarithmic plots of the time-domain evolution for three types of perturbations ($Q=0.3, 0.5, 0.8$) with $\alpha=0.5$ and $l=2$.}
	\label{f9}
\end{figure*}

In this paper, we set \(A=1\), \(a=20\), \(\sigma=\sqrt{2}\), place the observer at \(r_*^{\rm obs}=150\), and use a step size of \(\Delta=1/10\). The time-domain profiles for the three types of perturbations (\(l=0, 1, 2\)) are shown in Fig.~\ref{f7}. The profiles for scalar perturbations (\(\alpha=0.2, 0.3, 0.5\)) are shown in Fig.~\ref{f8}. And the profiles for the three types of perturbations (\(Q=0.3, 0.5, 0.8\)) are shown in Fig. \ref{f9}.

From the time-domain profile, we can clearly observe three stages of time-domain evolution: the initial perturbation stage, the ringdown stage (or quasinormal modes stage), and the late-time power-law tail. The ringdown stage is also known as the quasinormal modes stage. The analysis in Sec.~\ref{chapter4} of this paper focuses on the frequency-domain characteristics of this stage. While frequency-domain analysis yields complex-valued information at discrete frequencies, time-domain evolution naturally reveals the entire process of the perturbation transitioning from its initial state to its final state. Next, we analyze the influence of parameters on the system.

As shown in Fig.~\ref{f7}, for fixed modified gravity parameters $\alpha$ and charge $Q$, we find that the time-domain evolution for $l=0$ is significant only for the scalar field. The electromagnetic and gravitational fields do not exhibit complete time-domain waveforms, indicating that standard quasinormal mode oscillations are not excited. Regarding the time-domain profiles, the study of quasinormal modes requires selecting the stage characterized by a complete waveform, specifically the ringdown phase. Consequently, studies typically focus on $l \geq 1$ for electromagnetic fields and $l \geq 2$ for gravitational fields. During the black hole ringdown phase, perturbations decay via quasinormal modes oscillations, with the duration determined by the damping rate $|\omega_I|$. Observations of the time-domain profiles reveal that gravitational perturbations decay the slowest, resulting in the longest ringdown signal duration. Scalar perturbations decay the fastest, yielding the shortest ringdown, and electromagnetic perturbations fall in between. These findings are consistent with the results obtained in Sec.~\ref{chapter4} of this paper.

In Fig.~\ref{f8}, we study the impact of the modified gravity parameter $\alpha$ on the time domain evolution under scalar perturbation. As the modified gravity parameter $\alpha$ increases, the duration of the ringdown signal becomes longer, and as the angular quantum number $l$ increases, the decay becomes slower. It can be concluded that correcting the modified gravity parameter $\alpha$ and the angular quantum number $l$ will prolong the duration of the ringdown signal, affect the time when the black hole enters the power law tail, and have only a slight impact on the time domain waveform in the initial perturbations stage. Finally, we investigate the influence of the charge $Q$ on the time-domain evolution, as shown in Fig.~\ref{f9}. As the charge $Q$ increases, the decay becomes faster and the system enters the tail phase earlier. That is, the imaginary part of the quasinormal frequency decreases, while its magnitude increases, which is consistent with the result obtained by the frequency-domain analysis. The ordering of the decay rates still follows the pattern: gravitational perturbations decay the slowest, electromagnetic perturbations are intermediate, and scalar perturbations decay the fastest.

\section{Conclusion}\label{chapter6}
Within the framework of Scalar-Tensor-Vector Gravity (STVG), this paper systematically investigates the quasinormal modes and time-domain evolution of Reissner-Nordström MOG black holes perturbed by massless scalar, electromagnetic, and gravitational fields. Quasinormal frequencies were calculated using the 3rd- to 6th-order WKB method extended to the 13th order via Padé approximation, while the complete evolution of the perturbation fields in the time domain was simulated using the finite-difference method.

We investigated the quasinormal modes of Reissner-Nordström MOG black holes using both frequency-domain and time-domain methods. In the frequency-domain analysis, we employed the WKB method and Padé approximants to calculate the quasinormal frequencies for three types of perturbations for various parameter values. The Padé-improved WKB method enhanced computational accuracy and enabled the calculation of higher-order approximations, thereby overcoming the limitations of the standard WKB approach. For the time-domain analysis, we utilized the finite-difference method to numerically evolve the waveforms resulting from external field perturbations. This approach complements the frequency-domain analysis by capturing the initial perturbation and power-law tail phases, allowing for a more intuitive study of the perturbation behavior.

The quasinormal modes of Reissner-Nordström MOG black holes are influenced by the modified gravity parameter $\alpha$; an increase in $\alpha$ suppresses oscillation frequencies and reduces damping for all three types of perturbations, distinguishing them from those of standard Reissner-Nordström black holes. The parameter $\alpha$ affects the duration of the ringdown phase and the actual oscillation frequency, thereby influencing the onset of the power-law tail, though its impact on the initial perturbation phase is minimal. The prolonged duration of the gravitational signal makes it the most promising channel for extracting quasinormal modes (QNM) information from gravitational-wave observations and offers an optimal window for testing modified gravity theories using ringdown signals.

The time-domain waveforms obtained via the finite-difference method fully illustrate the entire evolutionary process, ranging from the initial perturbation and quasinormal oscillations to the late-time power-law tail. The oscillations and decay observed in the time domain show qualitative agreement with the real and imaginary parts of the WKB frequencies, thereby validating the reliability of both methods. Time-domain analysis further reveals that increasing $\alpha$ not only prolongs the ringdown phase but also influences the power-law decay behavior of the late-time tail to some extent. Although the dominant power-law index remains unchanged in asymptotically flat spacetimes, the MOG parameter alters the contribution of sub-dominant terms, leaving an observable imprint on the decay during the intermediate late-time regime.

These findings have important implications for the gravitational-wave signatures of black holes in MOG. If nature is described by MOG, the ringdown signal following a black hole merger would persist longer than predicted by General Relativity. This would directly impact the construction of matched-filtering templates and the estimation of signal-to-noise ratios in gravitational-wave data analysis. In summary, through a systematic analysis combining frequency-domain and time-domain approaches, this paper reveals the quantitative influence of MOG parameters on the perturbation response of charged black holes, providing a theoretical basis and numerical reference for testing modified gravity theories via black hole spectroscopy.

\begin{acknowledgments}
We thank Emanuele Berti and Caio F. B. Macedo for their helpful discussions. The work is supported by Yunnan Xingdian Talent Support Program - Young Talent Project.	
\end{acknowledgments}

\bibliography{refs}

\end{document}